\RequirePackage{fix-cm}
\documentclass[twocolumn,epjc3]{svjour3}  
\smartqed  
\RequirePackage{color,graphicx,url}
\usepackage{amsfonts,amsmath,amssymb,bm,xspace}
\usepackage{lipsum}
\usepackage[T1]{fontenc} 
\usepackage{epsfig}  
\usepackage{graphicx}  
\usepackage{amssymb}
\usepackage{amsmath}
\usepackage{pifont}
\usepackage[english]{babel}
\usepackage{color}  
\usepackage{multirow}
\definecolor{darkgreen}{rgb}{0.2,0.5, 0.2}
\usepackage{lipsum}
\usepackage{mathtools}
\usepackage{cuted}
\usepackage{dcolumn,booktabs}
\usepackage{ulem}
\newcolumntype{d}[1]{D{.}{.}{#1}}
\usepackage{cite} 
\usepackage{soul}
\usepackage{xcolor}
\newcommand{\sat}{\mathrm{sat}}

\journalname{Eur. Phys. J. A}
\begin{document}

\title{Skyrme-Hartree-Fock-Bogoliubov mass models on a 3D mesh: III. From atomic nuclei to neutron stars}


\author{Guilherme Grams\thanksref{e1,addr1}
        \and
        Wouter Ryssens\thanksref{addr1} 
        \and
        Guillaume Scamps\thanksref{addr2} 
        \and
        Stephane Goriely\thanksref{addr1} 
        \and
        Nicolas Chamel\thanksref{addr1} 
}
\thankstext{e1}{e-mail: guilherme.grams@ulb.be}

\institute{Institut d’Astronomie et d’Astrophysique, Université Libre de Bruxelles, Brussels, Belgium \label{addr1}
           \and
           Department of Physics, University of Washington, Seattle, USA. \label{addr2}
}

\date{Received: date / Accepted: date}
\maketitle

\begin{abstract}
We present BSkG3, the latest entry in the Brussels-Skyrme-on-a-grid series of large-scale models of nuclear structure based on an energy density functional. Compared to its predecessors, the new model offers a more realistic description of nucleonic matter at the extreme densities relevant to neutron stars. 
 This achievement is made possible by incorporating
a constraint on the infinite nuclear matter properties at high densities in the parameter adjustment, ensuring in this way that the predictions of BSkG3 for the nuclear Equation of State are compatible with the observational evidence for heavy pulsars with $M > 2 M_{\odot}$. Instead of the usual phenomenological pairing terms, we also employ a more microscopically founded treatment of nucleon pairing, resulting in extrapolations to high densities that are in line with the predictions of advanced many-body methods and are hence more suited to the study of superfluidity in neutron stars. By adopting an extended form of the Skyrme functional, we are able to reconcile the description of matter at high densities and at saturation density: the new model further refines the description of atomic nuclei offered by its predecessors. A qualitative improvement is our inclusion of ground state reflection asymmetry, in addition to the spontaneous breaking of rotational, axial, and time-reversal symmetry. Quantitatively, the model offers lowered root-mean-square deviations on 2457 masses (0.631 MeV), 810 charge radii (0.0237 fm) and an unmatched accuracy with respect to 45 primary fission barriers of actinide nuclei (0.33 MeV). Reconciling the complexity of neutron stars with those of atomic nuclei establishes BSkG3 as a tool of choice for applications to nuclear structure, the nuclear equation of state and nuclear astrophysics in general.
\end{abstract}
%

%

\section{Introduction}
\label{intro}

Nucleons in limited numbers can form self-bound composite objects: atomic 
nuclei composed of up to 238 nucleons occur naturally on earth while heavier
isotopes, up to $^{294}$Og so far, can be synthesised in accelerators~\cite{Oganessian06}. 
The properties of nuclei are directly relevant to astrophysics: nuclear 
reactions and decays produce the energy that powers stars throughout their 
evolution~\cite{Eddington20}. Although reaction rates are strongly influenced
by temperature and extremely neutron-rich systems can be produced in violent 
phenomena, in most astrophysical conditions atomic nuclei remain the relevant
organisational unit of nucleons with central densities never deviating much 
from saturation $\rho_{\rm sat} \approx 0.16$ fm$^{-3}$.

Neutron stars (NSs), exotic and extremely compact objects created in the gravitational 
core-collapse of massive stars at the end of their lives are notable exceptions:
these stars contain about $10^{57}$ nucleons under extreme gravitational
pressure.
Their interior is stratified into distinct layers, each of which features a 
different arrangement of protons and neutrons. 
The surface of cold-catalyzed NSs is believed to be covered by an ocean 
of $^{56}$Fe and neighboring elements~\footnote{The case of accreting NSs is more 
complex: their surface also contains light elements such as hydrogen and helium transferred 
from the stellar companion, and heavier elements produced by thermonuclear 
explosions, see e.g. Ref.~\cite{blaschke2018} and references therein.}
while the layers beneath are solid and composed of 
highly exotic nuclei that become progressively more neutron-rich with increasing 
depth~\cite{Chamel08b}. The point at which neutrons start to drip out of nuclei 
delimits the boundary 
between the outer and inner crust where a liquid of free neutrons coexists with 
neutron-rich clusters containing hundreds of nucleons~\cite{Chamel08b}. The dissolution 
of the crust at about half saturation density marks the transition to the core, which 
consists of a liquid mixture of protons, neutrons, and leptons. The composition of the 
inner part of the core in the most massive NSs remains uncertain. 
Near the crust-core transition, some models predict 
the existence of a nuclear pasta mantle where neutrons and protons form exotic
structures such as long cylinders (`spaghetti') or plates (`lasagna') due to 
Coulomb frustration~\cite{Ravenhall83,Hashimoto84}. The Equation of State (EoS) of dense matter, 
{\it i.e.}, the thermodynamical relation between the mean energy density of matter 
and pressure, is the key microscopic ingredient to determine the macroscopic properties of a NS~\cite{Oertel17}.

Understanding the properties of matter
in all astrophysical environments is an enormous challenge. Even near saturation density, 
the difficulty of creating and handling radioactive isotopes implies that 
experimental efforts cannot possibly measure the properties of all relevant
isotopes and reactions. 
So far, the only earth-based method capable of investigations beyond saturation 
density relies on relativistic collisions of heavy ions to produce high-density
nucleonic matter that is out of equilibrium for most of its extremely short 
lifespan~\cite{danielewicz02,Sorensen23}.

Given these experimental difficulties, it falls to nuclear theory to build 
models that are capable of extrapolation to all relevant regimes. Ideally, a 
single model would provide us with a complete EoS, i.e. one that 
spans the enormous range from below saturation density to that prevailing in the core 
of NSs for arbitrary proton-neutron asymmetry, but also with all properties
of nuclear structure that are relevant to the modelling of nuclear reactions 
and decays, from binding energies to more complicated quantities such as
nuclear level densities, optical potentials, or strength functions~\cite{Arnould20}.
A complete description is particularly desirable to explore 
nucleosynthesis through the so-called rapid neutron capture 
process or r-process: when NSs collide, neutron-capture reactions and 
decays of different kinds compete to transform any ejected matter
into heavy elements. The combined observation of both the gravitational waves 
emitted by such an event and the electromagnetic radiation emitted by the ejected matter, 
the kilonova~\cite{Metzger10}, has recently confirmed NS mergers as a locus of
r-process nucleosynthesis~\cite{Kasen17}.

The complexity of the nucleon-nucleon interaction and the quantum many-body problem
renders a complete description of nucleonic matter a daunting task, so much
that all models rely on at least a few phenomenological ingredients. The 
dependence on the latter should however be minimized to gain a measure of 
confidence in our extrapolations to unknown regimes: it is essential to 
derive the emergent properties of nucleonic matter from the microscopic physics
of the nucleon-nucleon interaction as much as possible. Ab initio approaches arguably come closest
to this ideal but are limited (i) in nucleon number in the case of nuclei~\cite{Hergert20}
despite recent progress~\cite{Hu22} and (ii) to densities not much higher than 
saturation in the case of infinite homogeneous nuclear matter (INM)~\cite{Drischler19}.

Self-consistent models based on nuclear energy density functionals (EDFs) 
provide an alternative: they offer a quantum description of any nuclear system
in terms of its constituent neutrons and protons that is sufficiently tractable 
for application to the entirety of the nuclear chart and the different regions 
of NSs~\cite{Bender03}.
The key to the success of such models is their (comparatively) simple 
formulation in terms of the nucleonic densities connected to an effective 
nucleon-nucleon interaction, enabling both the accurate modeling of nuclear 
properties
at the basic mean-field level and applications to INM. Purely theoretical 
considerations are typically not sufficient to guide the construction of such 
an analytical form: in practice one adopts a phenomenological form with free 
parameters that are adjusted to nuclear data\footnote{The density matrix 
expansion of Negele and Vautherin~\cite{negele1972} and other similar 
techniques~\cite{carlsson2010} provide a way to derive an analytical 
form starting from a parameterisation of the bare nucleon-nucleon interaction.
This has not found widespread adoption for a variety of reasons but has for
instance been used in Refs.~\cite{Stoitsov10,Perez18} to derive an EDF from on 
chiral effective field theory, although not without free parameters.
}.
To ensure that predictions are as reliable as possible, those data should be
carefully selected and as comprehensive as possible. This includes measurements 
of a variety of properties of a wide range of nuclei, from stable to
exotic neutron-rich isotopes, but also information on the EoS up to a few
times saturation density as obtained from heavy-ion collisions or ab initio
calculations. Astrophysical observations of NSs of masses
above $2 M_{\odot}$ provide an additional constraint at even higher densities.

We have recently started the development of a new set of models based on EDFs
of the Skyrme type: the Brussels-Skyrme-on-a-Grid (BSkG) series~\cite{Scamps21}.
Like their predecessors, the BSk-family~\cite{Samyn02,Goriely16}, the BSkG models 
aim to provide a global description of nuclear structure that is as microscopic 
as possible, with the specific focus of providing input for astrophysical 
applications. Since differences of binding energies set the energy scale for 
all nuclear reactions and decays, the objective function for these models
included all known nuclear masses of nuclei with $N,Z\geq 8$. As a result, 
the BSkG1~\cite{Scamps21} and BSkG2~\cite{Ryssens22} models achieved a 
root-mean-square (rms) deviation with respect to 2457 known masses of 0.741 and 
0.678~MeV, respectively, which is competitive with more phenomenological models~\cite{Moller16}.
Most other EDF-based models are adjusted to a limited number of (often spherical) 
nuclei and typically do not reach this level of accuracy despite the 
wide variety of EDF forms and fitting protocols available in the literature;
Refs.~\cite{kortelainen2010,Nakada10a,kortelainen2012,kortelainen2014,Jodon16,gil2017,Bennaceur_2017,becker2017,Reinhard17,Bulgac18,Perez18,Bennaceur_2020,giuliani2022,zhao2022a,Baldo23,batail2023} 
form a (necessarily non-exhaustive) set of recent examples. On the other hand, dedicated mass models augmented with machine 
learning techniques can reach rms deviations lower than 0.1~MeV~\cite{Niu22} 
but such approaches lack the reach 
in terms of additional observables that EDF-based models can offer. 
The BSkG-series combines the reproduction of known masses with, among other things, 
an excellent global description of known charge radii, the systematics of 
nuclear deformation and the fission properties of actinide nuclei~\cite{Ryssens23b}.

Crucial to the success of all self-consistent EDF-based models is the concept of
spontaneous symmetry breaking: by allowing for deformed mean-field 
configurations that do not respect all symmetries of nature, such models can 
grasp a large part of the effect of nuclear collectivity\footnote{
Though we note that spontaneous symmetry breaking cannot fully account for all 
types of collective motion, even if completely symmetry-unrestricted global
calculations would be feasible. Global models tacitly assume that the model 
parameters can absorb most of the missing physics and/or include phenomenological 
corrections. We do both, see Sec.~\ref{sec:massmodel}. } while remaining at 
the mean-field level and thus keeping global calculations tractable. 
Large-scale EDF-based models have accounted for rotational symmetry breaking 
and the appearance of nuclear deformation for a few decades now~\cite{Goriely2001}, 
but have not moved beyond the assumption of axial symmetry before the advent 
of BSkG1, which was the first large-scale EDF-based model to consistently employ a 
three-dimensional representation of the nucleus, thereby exploring the 
physics of triaxial deformation on the scale of the nuclear chart~\cite{Scamps21}.
With BSkG2 we went even further, allowing odd-mass and odd-odd nuclei to 
spontaneously break time-reversal symmetry and benchmarking the influence 
of spin and current densities and the so-called `time-odd' terms in the Skyrme
EDF they contribute to on a global scale~\cite{Ryssens22}.

Despite the quality of their description of nuclei, BSkG1 and BSkG2 are not 
quite satisfactory for the study of NSs for two reasons. First is the 
incompatibility of their predictions with observations of NSs: 
BSkG1 and BSkG2 each produce EoSs that lead to a maximum NS 
mass of about $1.8 M_{\odot}$ and thus are incompatible with the 
existence of heavy pulsars whose mass exceeds $2 M_{\odot}$, 
such as J1614-2230 and J0740+662~\cite{Demorest2010,Riley21,Miller21}. 
A second deficiency is their treatment of pairing: like most Skyrme models,
BSkG1 and BSkG2 rely on a simple and entirely phenomenological ansatz for the 
pairing terms of the EDF. If the relevant parameters are constrained on the 
properties of finite nuclei, such an approach leads to unrealistically
large $^1$S$_0$ pairing gaps in pure neutron matter when compared to advanced many-body 
treatments of the latter. A realistic description of pairing is particularly 
relevant to the inner crust and outer core of NSs, with implications 
for various astrophysical phenomena~\cite{Chamel17,Sedrakian19,Nils21}. 

We present here BSkG3, a new entry in the BSkG-series that addresses these
limitations, and is thus much better suited to the study of NSs
than its predecessors. 
To achieve this, we rely on ideas that were already employed by some of the 
latest BSk-models: i) we use a more microscopically-grounded treatment of
nucleon pairing that reproduces the $^1$S$_0$ pairing gaps in INM 
as predicted by extended Brueckner-Hartree-Fock calculations~\cite{Cao06,Chamel08}, 
and ii) we force the model to accommodate the existence of heavy pulsars by 
constraining the EoS at high densities~\cite{Chamel09}. Such constraints are 
difficult to reconcile with a competitive global reproduction of masses when 
relying on a Skyrme EDF of the traditional form: for this reason, we move here 
to the extended form proposed in Refs.~\cite{Chamel09,Goriely16}. 

We also further leverage the power of symmetry breaking: we now allow nuclei 
to take reflection asymmetric shapes with finite octupole deformation. Combined 
with the breaking of rotational, axial and time-reversal symmetries, this makes
our calculations the most general global study of octupole deformation to date
and the first to systematically include odd-mass and odd-odd nuclei.
We find that static octupole deformation is only relevant to ground 
states in specific regions of the nuclear chart, confirming the 
expectations of earlier studies~\cite{cao2020,Robledo11,Agbemava16}.
The impact of this degree of freedom as measured by the global accuracy of our
model for masses and radii is thus limited, but it does strongly affect the 
binding energy of exotic neutron-rich nuclei near $N \sim 196$, possibly affecting 
the simulations of r-process nucleosynthesis. It also allows our model to 
connect to the experimental body of evidence of reflection asymmetry in nuclear
ground states such as the occurrence of rotational bands of alternating parity~\cite{Butler96}.

This paper is organized as follows: we first discuss the model ingredients of 
BSkG3 in Sec.~\ref{sec:massmodel}, focusing on the differences with preceding 
models of the BSk- and BSkG-series. Sec.~\ref{sec:optimization} details 
the adjustment of the model parameters. We report the final values of 
the latter in Sec.~\ref{sec:BSkG3} and discuss in detail the models performance
with respect to several properties of atomic nuclei. We study the INM predictions 
of the new model in Sec.~\ref{sec:INM} and link these to the structure of 
NS in Sec.~\ref{sec:NS}, reserving our conclusions and outlook 
for Sec.~\ref{sec:conclusions}.
We provide a list of abbreviations and acronyms we use in~\ref{app:acronyms}, 
    the complete definition of all EDF coupling constants in terms of 
    model parameters in~\ref{app:couplingconstants}, as well as
    the equations for the rotational, vibrational and Wigner energies in~\ref{app:corrEnergies} 
    and an explanation of the tables of our results that we provide as supplementary material in~\ref{app:explanation}.

\section{Ingredients of the mass model}
\label{sec:massmodel}

\subsection{The nuclear binding energy and atomic mass}

We represent an atomic nucleus with $N$ neutrons and $Z$ protons with a 
many-body state of the Bogoliubov type, whose total energy $E_{\rm tot}$ we define as:
\begin{align}
E_{\rm tot} &= E_{\rm HFB} + E_{\rm corr} \, . 
\label{eq:Etot}
\end{align}
We call $E_{\rm HFB}$ the mean-field energy and $E_{\rm corr}$ is a set
of corrections that account (approximately) for correlations that cannot be 
captured by single mean-field reference state constructed from separate
neutron and proton orbitals. More precisely, the mean-field energy consists of five parts:
\begin{align}
E_{\rm HFB} &= E_{\rm kin} + E_{\rm Sk} + E_{\rm pair} + E_{\rm Coul}  + E^{(1)}_{\rm cm}\, ,
\label{eq:Ehfb}
\end{align}
which are, respectively, the kinetic energy, Skyrme energy, pairing energy, Coulomb energy and
the one-body part of the centre-of-mass correction~\cite{Bender00}. The correction 
energy $E_{\rm corr}$ is written in terms of four parts:
\begin{align}
E_{\rm corr}&= E_{\rm rot} + E_{\rm vib} + E^{(2)}_{\rm cm} + E_{\rm W}  \, ,
\label{eq:Ecorr}
\end{align}
which are, respectively, the rotational correction, the vibrational correction, 
the two-body part of the 
centre-of-mass correction~\cite{Bender00}, and the Wigner energy~\cite{Goriely03}. 
We will also refer to $E_{\rm corr}$ as the collective correction. 
The total energy $E_{\rm tot}$ is minus the binding energy of the nucleus; the 
atomic mass $M(N,Z)$ of an atom composed of this nucleus and $Z$ electrons is then given by
\begin{equation}
M(N,Z) = E_{\rm tot} + N M_n + Z (M_p + M_e) - B_{\rm e}(Z) \, .
\label{eq:mass_def}
\end{equation}
In this equation, $M_{n/p}$ are the masses of free neutrons and protons, 
$M_e$ is the mass of an electron 
and $B_{\rm e}(Z)$ is a simple analytical estimate for the binding energy of 
the electrons~\cite{Lunney03,AME2020}.

Most model ingredients in Eqs.~\eqref{eq:Ehfb} and \eqref{eq:Ecorr} are identical
to those employed in the BSkG1 and BSkG2 models. There are however two exceptions 
compared to the preceding models: (i) we employ a more general form of the Skyrme 
EDF and (ii) we rely on a less phenomenological treatment of the pairing energy.
The rest of this section discusses these two differences in detail. For a more 
detailed description of all other model ingredients, we refer the interested 
reader to Refs.~\cite{Scamps21,Ryssens22}.

\subsubsection{The Skyrme energy}
\label{sec:sk}

As is standard practice, we write the Skyrme energy as an integral over  
four energy densities $\mathcal{E}_{t, \rm e/o}(\bold{r})$:
\begin{align}
\label{eq:Eskyrme}
E_{\rm Sk} &= \int d^3 \bold{r} \sum_{t = 0,1} \left[ 
                                            \mathcal{E}_{t, \rm e}(\bold{r})
                                          + \mathcal{E}_{t, \rm o} (\bold{r}) 
                                          \right]
                                          \, ,
\end{align}
where $t=0,1$ is an isospin index. Different terms in the energy densities are combinations of different local densities and their derivatives: 
we employ the set ($\rho_t(\bold{r}), \tau_t(\bold{r})$, $\bold{J}_t(\bold{r})$, 
$\bold{s}_r(\bold{r})$, $\bold{j}_r(\bold{r})$), whose definitions are standard
in the literature~\cite{Ryssens21}. The first three densities 
are even under time-reversal, while the latter two are odd. Since the energy 
densities themselves are necessarily time-even under time-reversal, their 
constituent terms can be separated in those bilinear in the time-even densities
($\mathcal{E}_{t, \rm e} (\bold{r})$) and those bilinear in the time-odd densities 
($\mathcal{E}_{t, \rm o} (\bold{r})$).

Seventeen model parameters, {\it i.e.} six pairs $(t_i,x_i)$ ($i=0, \ldots, 5$) together
with three exponents $\alpha, \beta$ and $\gamma$ and two spin-orbit parameters
$W_0$ and $W_0'$, specify sixteen coupling constants $\{C_t\}$, 
ten of which depend on the isoscalar density $\rho_0(\bold{r})$. Dropping the 
position-dependence of the latter to lighten notation for coupling constants, 
we employ the energy densities:
%
\begin{align}
\label{eq:Skyrme_te}
\mathcal{E}_{t, \rm e}(\bold{r})
& =  
              C^{\rho\rho}_t(\rho_0) \, \rho_t^2 (\bold{r})
            + C^{\rho\tau}_t  (\rho_0 )\, \rho_t (\bold{r}) \, \tau_t (\bold{r})  \nonumber \\
&           + C^{\rho \nabla  J}_t  \rho_t (\bold{r}) \, \boldsymbol{\nabla} \cdot \bold{J}_t (\bold{r}) \nonumber \\ 
&           + C^{\rho \Delta \rho}_t  \, \rho_t (\bold{r}) \, \Delta \rho_t (\bold{r})  \nonumber \\
&           + C^{\nabla \rho \nabla \rho}_t  (\rho_0 )\, \boldsymbol{\nabla}  \rho_t (\bold{r}) \cdot \boldsymbol{\nabla}  \rho_t (\bold{r})  \nonumber \\
&           + C^{\rho \nabla \rho \nabla \rho}_t  (\rho_0 )\, \rho_t (\bold{r}) \boldsymbol{\nabla} \rho_0 (\bold{r})  \cdot \boldsymbol{\nabla}  \rho_t (\bold{r}) \\
\mathcal{E}_{t, \rm o}(\bold{r})
& =  
                C^{s s}_t (\rho_0) \, \bold{s}_t (\bold{r}) \cdot \bold{s}_t (\bold{r})
              + C^{j j}_t (\rho_0 )\, \bold{j}_t (\bold{r}) \cdot \bold{j}_t (\bold{r})   \nonumber \\
&             + C^{j \nabla s}_t \, \bold{j}_t (\bold{r}) \cdot \nabla \times \bold{s}_t (\bold{r}) 
\, .
\label{eq:Skyrme_to}
\end{align}

Eqs.~\eqref{eq:Skyrme_te} and~\eqref{eq:Skyrme_to} differ from the energy 
densities of BSkG2: the final two terms in Eq.~\eqref{eq:Skyrme_te} 
did not appear in Ref.~\cite{Ryssens22} and the coupling constants
$C^{\rho \tau}_r(\rho_0)$ and $C^{jj}_t(\rho_0)$ were not density-dependent. 
This extended form of the Skyrme EDF arises naturally if one generalizes the terms involving 
$t_1$ and $t_2$ in the standard functional to depend on the density: the 
relevant new parameters are $t_4,x_4,t_5,x_5, \beta$ and $\gamma$. This 
generalization was already incorporated in the latest BSk models~\cite{Goriely16}
but is at the time of writing not widely spread.
 
In terms of the short-hands $C^{+/-}_{0t}(t,x)$ defined in \ref{app:couplingconstants}, 
we write for the time-even part of the EDF
%
\begin{alignat}{2}
%
\label{eq:Crhotau}
C^{\rho \tau}_t (\rho_0)        =& + \frac{1}{2} C^+_{0t}(t_1,x_1)  + \frac{1}{2} C^-_{0t}(t_2,x_2)                            \nonumber \\
                                 & + \frac{1}{2}C_{0t}^+(t_4,x_4)\rho_0^{\beta}  + \frac{1}{2}C_{0t}^-(t_5,x_5)\rho_0^{\gamma}  \, ,     \\  
%
\label{eq:Cnrhonrho}
C^{\nabla \rho\nabla \rho}_t (\rho_0) =& + \frac{3}{8} C^+_{0t}(t_4,x_4)\rho_0^{\beta}
                                         - \frac{1}{8} C^-_{0t}(t_5,x_5) \rho_0^{\gamma}  \, , \\
C^{\rho \nabla \rho\nabla \rho}_t (\rho_0) =& - \frac{1}{2} C^+_{0t}(t_4,x_4) \rho_0^{\beta-1} \, .
\label{eq:Crhonrhon}
\end{alignat} 
In the time-odd part, only the term $C^{jj}_t$ is affected by the new parameters:
the requirement of local gauge invariance imposes that $C^{jj}_t(\rho_0)=-C^{\rho \tau}_t (\rho_0)$~\cite{Raimondi_2011}. 
The expressions for all other coupling constants in terms of the model 
parameters $(t_i,x_i)$ ($i=0,\ldots,3$) are identical to those of BSkG2, 
we repeat them for completeness in \ref{app:couplingconstants}. However, 
we draw the attention of the reader to a change in notation compared to the 
previous BSkG models: here we associate the exponent $\gamma$ with the coupling 
constants in Eqs.~\eqref{eq:Crhotau}-\eqref{eq:Crhonrhon} while in 
Refs.~\cite{Scamps21,Ryssens22} the same symbol determined the density 
dependence of $C_t^{\rho \rho}(\rho_0)$. 

The form of the EDF we employ here is directly inspired by that of 
Ref.~\cite{Chamel09} and, as we will show, shares its chief advantage in that it
allows for a stiff EoS of neutron matter (NeutM) in combination with an 
excellent mass fit. However, there are a few differences. 
First, we do not include any term quadratic in the spin-current 
density $J_{\mu \nu}(\bold{r})$ nor its time-odd counterpart of the form
$\bold{s}_t(\bold{r}) \cdot \bold{T}_t(\bold{r})$. Second, we do not include
any term involving a gradient of the spin density $\bold{s}_t(\bold{r})$, as these
have a tendency to induce unphysical finite-size instabilities~\cite{Hellemans_2013}. 
We adhered to these policies for the standard Skyrme functional for the construction
of BSkG1 and BSkG2, and apply them here equally to the generalized EDF form of Ref.~\cite{Chamel09}.

Aside from these omitted terms, there are seemingly further differences between our 
formulation of the energy densities in Eqs.~(\ref{eq:Skyrme_te}--\ref{eq:Skyrme_to}) and their counterparts in Ref.~\cite{Chamel09}. 
These differences are of no consequence however: one can use partial integration
under the integral sign of Eq.~\eqref{eq:Eskyrme} to recombine the various
gradients in the energy densities in different ways, leading to many different 
expressions for $\mathcal{E}_{t, \rm e/o}(\bold{r})$ that are not identical yet 
physically equivalent. Eqs~\eqref{eq:Skyrme_te} and~\eqref{eq:Skyrme_to} 
are the expressions we found most convenient for numerical implementation.

\subsubsection{The pairing energy}
\label{sec:pairingEDF}

We supplement the Skyrme form in the particle-hole channel with the following density-dependent pairing functional:
\begin{align}
E_{\rm pair} &= \frac{1}{4}\sum_{ q=p,n}\int d^3   \bold{r} \, g_{q}(\rho_n, \rho_p) \tilde{\rho}_q^*(\bold{r})\tilde{\rho}_q(\bold{r})\, , 
\label{eq:pairing_energy}
\end{align}
where $\tilde{\rho}_q(\bold{r})$ is the pairing density of species $q=p,n$. 
Although the definition of this density is somewhat standard in the literature, 
see for example Ref.~\cite{Ryssens21}, it is not often mentioned that it does
not have a definite behaviour under time-reversal: its real part is time-even
while its complex part is time-odd~\cite{Ryssens22}. To avoid the ultra-violet
divergence of Eq.~\eqref{eq:pairing_energy}, we calculate the pairing density 
in the single-particle basis which diagonalises the single-particle hamiltonian
and weight the contribution of each state with single-particle energy $\epsilon$ 
to the pairing density with the following cutoff function:
\begin{align}
f_q &=  \bigg[ 1 + e^{(\epsilon - \lambda_q - E_{\rm cut})/\mu} \bigg]^{-1/4} \, ,
\label{eq:cutoff}
\end{align}
where $\lambda_q$ is the Fermi energy of species $q$, 
$E_{\rm cut}$ is an adjustable parameter of the model and we fix $\mu = 0.5$ MeV.
Note that Eq.~\eqref{eq:cutoff} eliminates the contribution from states at high 
energy but does not significantly affect states below the Fermi energy; note 
that this differs from our choice of cutoff for the preceding models. 

Both BSkG1 and BSkG2 rely on a widely-used empirical form for the function 
$g_q(\rho_n, \rho_p)$:
\begin{align}
g_q(\rho_n, \rho_p) &= V_{\pi q} \left[ 1 - \eta \left( \frac{\rho_0(\bold{r})}{\rho_{\rm ref}} \right)^{\xi}\right] \, ,
\label{eq:pairing_strength_pheno}
\end{align}
%
where $\eta, \xi, V_{\pi p}$ and $V_{\pi,n}$ are adjustable parameters\footnote{Refs.~\cite{Scamps21,Ryssens22} employ the symbol $\alpha$ 
instead of $\xi$ for the exponent in Eq.~\eqref{eq:pairing_strength_pheno}. We change
notation here to avoid confusion with the exponents in the particle-hole channel 
of the EDF.}
and the reference density is typically fixed at $\rho_{\rm ref} = 0.16$ fm$^{-3}$. 
Despite its success for pairing-related quantities
in nuclei, this recipe is not suited to NSs applications for two reasons. 
First, assuming $V_{\pi q}$ is negative, Eq.~\eqref{eq:pairing_strength_pheno} 
describes a transition to a regime of repulsive pairing for 
$\rho_0(\bold{r}) \geq \rho_{\rm ref}/\eta^{\alpha}$. The parameter values of
BSkG1 and BSkG2 imply this unphysical transition happens at $0.258$ and $0.396$
fm$^{-3}$, i.e. in density regimes that exist in low- and medium-mass NSs.

\begin{figure}
    \centering
    \includegraphics[width=.49\textwidth]{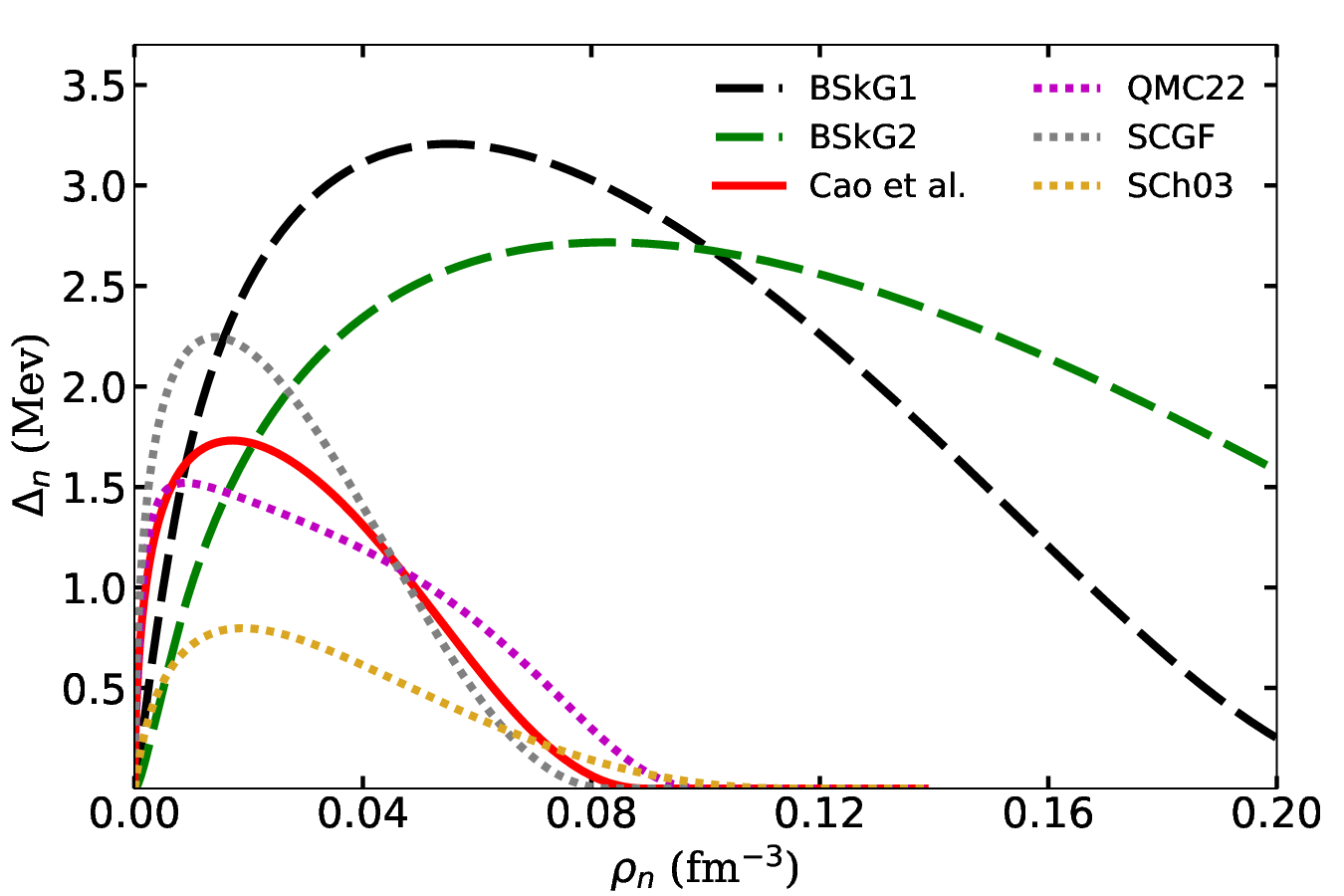}
    \caption{$^1$S$_0$ pairing gaps $\Delta_n$ in NeutM as a function of the neutron density $\rho_n$: 
    from Cao et al.~\cite{Cao06} (red full line), QMC22~\cite{gandolfi22} (magenta dotted line), SCGF~\cite{drissi22} (gray dotted line), SCh03~\cite{schwenk03} (yellow dotted line), 
    MB23~\cite{Krotscheck23} (brown dotted line), BSkG1 (black dashed line) and BSkG2 (green dashed line).
    The curves for BSk31 and BSkG3 are equal to the results of Cao et al. by construction. 
    }
    \label{fig:INM_gaps}
\end{figure}

The second, less obvious, reason is that when the parameters of empirical forms 
 such as Eq.~\eqref{eq:pairing_strength_pheno} are adjusted to the properties
 of nuclei, the resulting pairing gaps in INM are qualitatively different from the predictions of more
 advanced many-body methods. To illustrate this point, we show in Fig.~\ref{fig:INM_gaps} 
 the neutron $^1$S$_0$ pairing gaps $\Delta_{n}$ in NeutM as a function of the density 
 $\rho_n$ for the BSkG1 
 and BSkG2 
 models and compare them to the predictions of several references: the extended 
 Brueckner-Hartree-Fock (BHF) calculations of Ref.~\cite{Cao06}, 
 the calculations based on renormalization group theory (SCh03) of Ref.~\cite{schwenk03} 
 and more recent results obtained with Quantum Monte Carlo techniques (QMC22)~\cite{gandolfi22}, 
 self-consistent Green's functions (SCGF)~\cite{drissi22} 
 and a diagrammatic method (MB23)~\cite{Krotscheck23}. 
 Although there is a certain spread in the predictions of the advanced
 many-body calculations, it is immediately obvious that the predictions of BSkG1
 and BSkG2 largely fall outside this spread. The empirical pairing terms employed by 
 both BSkG1 and BSkG2 (as well as most of the standard volume and/or surface pairing interactions used in the framework of Skyrme-HFB calculations)
 result in large pairing gaps up to 3 MeV that persist up to high densities 
 while more microscopic approaches produce gaps that are generally smaller than 
 2 MeV and quickly decay beyond $\rho_n \sim 0.05$ fm$^{-3}$.

To address these deficiencies, we discard the phenomenological ansatz for 
$g_{q}(\rho_n, \rho_p)$ and rely instead on the approach of Refs.~\cite{Chamel08,chamel2010}:
it allows our new model to reproduce exactly any given set of pairing 
gaps in INM, $\Delta_q^{\rm INM}(\rho_n, \rho_p)$. We repeat the choice made
for BSk31 \cite{Goriely16} and take as starting point the INM pairing gaps of Ref.~\cite{Cao06}, 
thereby essentially forcing BSkG3 to reproduce the solid red curve in 
Fig.~\ref{fig:INM_gaps}. Ref.~\cite{Cao06} only provides gap for symmetric 
($\Delta^{\rm INM}_{\rm sm}$) and pure neutron matter ($\Delta^{\rm INM}_{\rm nm}$), 
which forces us to adopt an interpolation recipe to obtain the gap at arbitrary 
asymmetry $\delta$:
\begin{align}
\Delta_q^{\rm INM}(\rho_n, \rho_p) &= (1 - |\delta|) \Delta^{\rm INM}_{\rm sm}(\rho_0) + |\delta| \Delta_{\rm nm}(\rho_q) \, . 
\label{eq:interpolation_INM}
\end{align}
This linear interpolation is entirely empirical due to the lack of guidance from advanced many-body approaches, but guarantees positive gaps for arbitrary $\delta$.

We now start from the following ansatz for the function $g_{q}$:
\begin{align}
g_{q}(\rho_n, \rho_p) &=
V_{q}(\rho_n, \rho_p)  \left[
1 + \kappa_q (\nabla \rho_0)^2 \right]
\, , 
\label{eq:micro_strength}
\end{align}
where the $\kappa_{q}$ are adjustable parameters and $V_{q}(\rho_n, \rho_p)$ are 
the ``pairing strengths'' which reproduce the (interpolated) INM pairing 
gaps (Eq.~\ref{eq:interpolation_INM}). 
These strengths can be calculated via~\cite{Chamel08,chamel2010}:
\begin{align}
V_{q}(\rho_n, \rho_p) = - \frac{8 \pi^2}{I_q(\rho_n, \rho_p)} \left(\frac{\hbar^2}{2M^*_q(\rho_n, \rho_p)}\right)^{3/2} \, ,
\label{eq:Vq}
\end{align}
where
\begin{align}
I_q &= \int_{0}^{\lambda^{\rm INM}_q + E_{\rm cut}} d \xi \frac{\sqrt{\xi}}{\sqrt{(\xi- \mu_q)^2 + [\Delta^{\rm INM}_{q}(\rho_n, \rho_p)] ^2}} \, .
\label{eq:pairing_integral}
\end{align}
%
In Eq.~\eqref{eq:Vq}, $M_q^*(\rho_n, \rho_p)$ is the position-dependent effective mass of species $q$ defined through
\begin{align}
\frac{\hbar^2}{2M^*_q(\rho_n, \rho_p)} &=  \frac{\hbar^2}{2m_q} + \sum_{t=0,1} \frac{\partial \mathcal{E}_{t, \rm e}}{\partial \tau_q} \, .
\label{eq:effective_mass}
\end{align}
We calculate the functional derivative in Eq.~\eqref{eq:effective_mass} consistently 
from the Skyrme EDF, that is to say from the equations presented in Sec.~\ref{sec:sk}
\footnote{We take the established practice of modelling 
neutrons and protons with equal masses, i.e.~our calculations consistently 
use $\hbar^2/2m_q = 20.73553$ MeV fm$^2$ for both nucleon species, 
including in Eq.~\eqref{eq:effective_mass}. }. 
The integral in Eq.~\eqref{eq:pairing_integral} depends on the Fermi energy 
in INM $\lambda^{\rm INM}_q$, which we estimate using the Fermi wave-number~\cite{chamel2010}: 
\begin{align}
\lambda^{\rm INM}_q &\approx \frac{\hbar^2 k_{F,q}}{2 M_q^*} = \frac{\hbar^2}{2 M^*_q} (3\pi^2 \rho_q)^{1/3} \, .
\end{align}
%
%
For our three-dimensional representation, the numerical integration of 
Eq.~\eqref{eq:pairing_integral} for each mesh point would become costly. To
avoid this, we employ the analytical approximation for $I_q$ of Ref.~\cite{chamel2010} 
that was also used for some of the later entries in the BSk family~\cite{Goriely16}.

The parameters $\kappa_q$ play no role in all of these considerations since 
all gradients of the density vanish identically in INM. They are nevertheless
crucial to the overall success of our model: without the $\kappa_q$
we would likely not have been able 
to reproduce the pairing properties of finite nuclei~\cite{Goriely16} 
from the realistic pairing gaps in INM (as already inferred from Fig.~\ref{fig:INM_gaps}). 
The reason is that all advanced many-body calculations 
produce pairing gaps that are small for neutron densities that are encountered near
the surface of finite nuclei and essentially vanishes for densities typically 
found near their centre, $\rho \sim 0.08 $ fm$^{-3}$. 
While the introduction of the $\kappa_q$ parameters is phenomenological, 
the gradient terms in the pairing EDF are not without physical 
motivation: it has long been known that the coupling to collective surface 
vibration modes leads to an induced pairing-like interaction between 
nucleons~\cite{Idini16}, which contributes significantly to mass 
differences such as $\Delta^{(5)}_{n/p}$ as we will discuss below~\cite{Barranco99}. 
Large-scale models relying on a basic mean-field level description of the nucleus
cannot directly capture such correlations and thus typically do not distinguish 
between the bare and induced pairing interaction, especially when attempting to 
model the effect of both with a simple ansatz such as Eq.~\eqref{eq:pairing_strength_pheno}.
Since we aim at deriving here a realistic description of the pairing gaps in 
INM where surface vibrations are absent, it becomes natural to mock-up the 
effect of the induced interaction in finite nuclei with the simple gradient 
term in Eq.~\eqref{eq:micro_strength}. The physics of surface vibrations is
however not without interest for NSs: in the inner crust superfluid 
neutrons coexist with nuclear clusters where both the bare and induced pairing
interaction contribute. 

The use of gradient terms in the pairing channel to mock up such effects was,
to the best of our knowledge, first proposed in Ref.~\cite{fayans1996}, incorporated later in what is currently known as the Fayans EDF~\cite{Fayans94,Fayans00} and investigated in combination 
with a standard Skyrme EDF in Ref.~\cite{Reinhard17}. The goal of these 
studies differed from ours: they all aim at an improved reproduction of the 
odd-even staggering of nuclear charge radii for isotopic chains of spherical 
nuclei, exploiting the contribution of the neutron pairing terms in 
Eq.~\eqref{eq:micro_strength} to the proton mean-field potential. A detailed 
investigation of these terms in view of the reproduction of charge radii 
during the adjustment of a global model like ours is beyond the scope of this manuscript: we drop all contributions
of the pairing terms to the nuclear mean fields, as we did
for BSkG1 and BSkG2, see also Sec.~\ref{sec:BSk_comparison}.


In summary, the chief difference with the BSkG1 and BSkG2 models is the form of the 
function $g_q(\rho_n, \rho_p)$: in Refs.~\cite{Scamps21,Ryssens22} we took 
the simple and entirely phenomenological standard form corresponding to a mix between 
so-called ``surface'' and ``volume'' pairing. A more technical difference
concerns the treatment of the cutoff function: to stay close to the formalism
in INM, Eq.~\eqref{eq:cutoff} only cuts states of high energy, while
we employed cutoffs both above and below the Fermi energy for the
construction of BSkG1 and BSkG2. 

\subsection{Numerical set-up, symmetries, shapes and blocking}
\label{sec:numerics}

As for the construction of BSkG1 and BSkG2, we rely on the MOCCa code of Ref.~\cite{RyssensThesis}
to solve the self-consis\-tent Skyrme-HFB equations quickly and robustly~\cite{Ryssens19}.  
This tool iterates $N_N$ single-neutron and $N_Z$ single-proton wavefunctions on a 
three-dimensional cubic mesh with $N_{x/y/z}$ points in each Cartesian direction, 
equally spaced at a distance $dx$.
Through the use of Lagrange-mesh 
techniques~\cite{Baye86}, a modest choice of mesh parameters already leads to an
excellent numerical accuracy that is essentially independent of the shape of the
nucleus~\cite{Ryssens15}. We fixed $dx=0.8$ fm in all calculations to guarantee
a numerical accuracy on absolute energies on the order of 100 keV~\cite{Ryssens15}, 
but we chose slightly different numbers of mesh points and single-particle wavefunctions
depending on the context, as we will describe below.

It turned out that numerical safeguards are crucial in order to solve 
the self-consistent equations for the form of the Skyrme functional we employ 
here. The density-dependent terms in Eqs.~\eqref{eq:Skyrme_te}-\eqref{eq:Skyrme_to} give 
rise to contributions to the nuclear mean-field that are proportional to 
$\rho_0(\bold{r})$ raised to the power $(\alpha-1),(\beta-1),(\gamma-1)$ and even
$(\beta-2)$. Even though $\rho_0(\bold{r})$ is formally guaranteed
to be strictly positive everywhere, our values of $\alpha,\beta$ and $\gamma$ 
are smaller than one and typically lead to numerical problems in regions
of the simulation volume where $\rho_0(\bold{r})$ is small because of
the finite precision of floating point arithmetic. For this reason, we replace
all instances of $\rho^\nu_0(\bold{r})$ with $\nu < 0$ 
in the mean fields by $(\rho_0(\bold{r}) + \epsilon)^{\nu}$, 
where $\epsilon =  10^{-8}$ fm$^{-3}$
\footnote{In fact, we also employed this recipe for BSkG1 and BSkG2 with $\epsilon = 10^{-20}$ fm$^{-3}$. 
The additional density dependencies we employ here force us to enlarge the value of
$\epsilon$ significantly.}.

Another important aspect of our calculations is the self-consistent symmetries
imposed on the nuclear configurations: each additional symmetry assumption greatly 
reduces the computational burden but also removes degrees of freedom that 
can be relevant to the description of the nucleus. In all calculations, we 
assume the nuclear configurations to be invariant under the $z$-signature 
$\hat{R}_z$ and $y$-time-simplex $\check{S}^T_y$ operators~\cite{Dobaczewski00}, 
such that we were only required to numerically represent one-fourth of the 
entire simulation volume corresponding to the positive $x$- and $y$-axes.
Calculations under these conditions are however very demanding even for modest
mesh parameters such that, when possible, we further simplified calculations 
by imposing reflection symmetry, time-reversal symmetry or both 
as we elaborate below in more detail. 

Symmetry assumptions restrict the range of nuclear shapes that can be explored.
One way to characterize the latter is the multipole moments 
$\beta_{\ell m}$ of the nuclear density, that we define exactly as in Ref.~\cite{Scamps21}
for integer $\ell \geq 1$ and integer $0 \leq m \leq \ell $\footnote{We remind the reader that 
such multipole moments characterize the deformation of the nuclear volume and 
are for that reason not directly comparable to the deformation parameters
typically used in analytical parameterisations of the nuclear density, see the
discussion in Ref.~\cite{Ryssens23b}.}. 
In our most general calculations, all moments $\beta_{\ell m}$ for arbitrary $\ell$ and even $m$ are allowed to be non-zero.
In particular, this includes two quadrupole deformations 
$\beta_{20}$ and $\beta_{22}$ and two octupole deformations $\beta_{30}$ and 
$\beta_{32}$. For plotting 
and interpretation purposes, we will also use an equivalent, widely-used, 
characterization of the quadrupole deformation\footnote{It is entirely possible 
to similarly recast the octupole deformations $\beta_{30}, \beta_{32}$ in terms of an overall size of 
the octupole deformation $\beta_3$ and an associated angle $\gamma_3$~\cite{Hamamoto91},
but these variables are not widely used. } in terms of $(\beta_2, \gamma)$:
\begin{subequations}
\begin{align}
\label{eq:beta}
\beta_2  &= \sqrt{\beta_{20}^2 + 2 \beta_{22}^2 } \, , \\
\gamma &= \text{atan} \left( \sqrt{2}\beta_{22}/ \beta_{20} \right) \, . 
\label{eq:gamma}
\end{align}
\end{subequations}
These values characterize the shape of a nucleus and should not to be confused 
with the exponents of the density-dependent terms in the EDF. 
Although multipole moments of order $\ell \geq 4$ are less often discussed, they
generally do not vanish~\cite{Scamps21} and can impact both low- and 
high-energy experiments~\cite{Bemis73,Zumbro86,Ryssens23b}.

All multipole moments that are not restricted by symmetry are naturally included
in the self-consistent optimization of the mean-field energy $E_{\rm HFB}$.
The correction energy $E_{\rm corr}$ is however a complicated function that 
cannot easily be included self-consistently in the Skyrme-HFB equations: as 
for the previous BSkG models, we include it perturbatively in any given 
MOCCa calculation but semivariationally optimize the total energy $E_{\rm tot}$
by performing multiple calculations constrained to different values of 
$(\beta_{20}, \beta_{22})$~\cite{Scamps21,Ryssens22} to find the 
nuclear ground state. It would have been prohibitively expensive
to extend this two-dimensional search to include $\beta_{30}$ and/or $\beta_{32}$. 
Instead, we perform a semivariational search in $(\beta_{20}, \beta_{22})$
twice for each nucleus: once with the assumption of reflection symmetry and 
once without, forcing the nucleus to explore reflection asymmetric degrees of
freedom by activating a constraint on a finite value of $\beta_{30}$ for a few 
iterations. These searches yield two nuclear configurations (that 
coincide for many nuclei), from which we select the one with the lowest total 
energy as our calculated nuclear ground state.

Our semivariational procedure is not completely general: we restricted our 
calculations to $\gamma \in [0^{\circ}, 60^{\circ}]$. The simplest of our 
calculations respect reflection symmetry and time-reversal symmetry; in 
this case values of $\gamma > 60^{\circ}$ reflect different orientations of the 
nuclear configuration in the simulation volume. Breaking time-reversal symmetry 
allows the nucleus to develop a finite angular momentum while breaking reflection
symmetry allows for finite octupole deformation; both of these quantities define
a preferred direction in space such that the degeneracy due to the reorientation
of the nucleus is partially lifted. We ignore this complexity here entirely 
for computational reasons, but our restriction to $\gamma \in [0^{\circ}, 60^{\circ}]$
is not without physical motivation: (i) reorientation effects due to the presence
of angular momenta are small at least for odd-mass nuclei~\cite{Schunck10}
and (ii) our search covers the special case of shapes with $\gamma = 0^{\circ}$ 
and finite values of $\beta_{30}$ that is the subject of virtually all literature
on ground state octupole deformation. Nuclear shapes that combine non-axial
quadrupole deformation and octupole deformation are relevant to fission~\cite{Ryssens23}, 
but we are not aware of any indication that such shapes are relevant to
nuclear ground states.

In typical conditions, the Bogoliubov many-body state with the lowest energy for 
a fixed configuration of the nuclear mean fields will have even number parity
for both protons and neutrons. To correctly describe the physics of the 
unpaired nucleons, we excite one or two quasiparticle excitations with respect
to this lowest state when describing odd-mass or odd-odd nuclei, respectively. 
We employ the gradient solver strategy briefly described in 
Ref.~\cite{Ryssens23} to facilitate these calculations 
and to guarantee we obtain the state of lowest energy at convergence.
To limit the computational effort, we consider only quasi-particle excitations
with $z$-signature quantum number $+i$ which slightly limits the generality of
our calculations for odd-odd nuclei but is sufficient for odd-mass systems~\cite{Ryssens23}.
As for BSkG1 and BSkG2, allowing for triaxial deformation implies that we cannot
cleanly associate rotational quantum numbers with the ground-states of odd-mass
and odd-odd nuclei. When breaking reflection symmetry, we can no longer 
associate a parity quantum number with these states either. Remedying these
defects of our model would require very expensive symmetry-restoration techniques,
whose systematic application to large numbers of nuclei is not feasible today.

Having clarified these general aspects of our approach, we are now in a position 
to describe in detail the three types of calculations we performed during this 
work: (i) calculations targeting ground states during the parameter adjustment, 
(ii) calculations targeting ground states to build the final table of results, 
after the parameter adjustment and (iii) calculations targeting fission barriers. 
\paragraph{Ground state calculations during the fit}
To save on computational resources, we took $N_x = N_y = N_z = 32$ and iterated
only $N_N = N + 160$ and $N_Z = Z + 100$ single-particle states during the fit 
for a nucleus with $N$ neutrons and $Z$ protons~\cite{Scamps21}. 
For even-even nuclei we assume time-reversal but we did not do so for 
odd-mass or odd-odd systems. We impose reflection symmetry 
for all calculations, which allowed us to further reduce the numerical 
representation to just one-eight of the entire simulation volume. We emphasize
that we did not include the possibility of octupole deformation in the 
parameter adjustment, which is justified a posteriori by the limited
number of nuclei affected.

\paragraph{Final ground state calculations:}
To make sure the final results are well converged with respect to all numerical
parameters, our final calculations employed $N_x = N_y = N_z = 36$ mesh points in 
each Cartesian direction and $N_N = N + 400$ and $N_Z = Z + 240$ single-particle states. 
As during the parameter adjustment, we only assume time-reversal invariance 
for even-even nuclei. We did not assume reflection symmetry, giving a priori
all nuclei the freedom to use octupole deformation to lower their total energy
according to our search strategy explained above.


\paragraph{Fission calculations:}
To obtain the static fission properties of the actinide nuclei included in the 
fit protocol, see Sec.~\ref{sec:optimization}, we followed the procedure detailed
in Ref.~\cite{Ryssens23}; we repeat here only a few essential points. We take
the two quadrupole deformations as collective variables and construct two-dimensional
potential energy surfaces (PES) through repeated calculations constrained to
different values of $(\beta_{20}, \beta_{22})$. In each of these calculations, 
we take $N_x = N_y = 32$, $N_z = 40$ mesh points to accommodate the elongated shapes 
relevant to fission and iterate only $N_N = 440$ and $N_Z = 260$ states; these
choices certainly limit our accuracy with respect to the total energy but still allow
for a numerical accuracy of roughly 100 keV for energy differences such as the 
barriers and isomer excitation energies. We established in Ref.~\cite{Ryssens23} 
that the effect of time-reversal symmetry breaking on such static fission properties is small 
and, for this reason, assume conservation of time-reversal symmetry in all 
fission calculations, employing the equal-filling-approximation in the case of odd-mass and odd-odd 
nuclei to account for the blocking effect of odd nucleon(s)~\cite{Perez08}. 
To reduce the computational burden, we assume reflection symmetry for nuclear configurations with $\beta_{20} < 1$. We relax this restriction at larger deformations, as the actinide nuclei we consider here typically exploit octupole deformation to lower their energy at deformation above $\beta_{20} \approx 1$\cite{Ryssens23}.
From the PES thus constructed, we used a flooding model to determine the lowest 
energy fission path (LEP) connecting the ground state with a fissioned system. 
We obtain through interpolation the excitation energy of the local maxima 
along this path, which by construction are saddle points on the two-dimensional 
PES. It is these excitation energies that we compare to the empirical values
listed in the RIPL-3 database~\cite{Capote09}. A final remark concerns nomenclature: it is 
natural to discuss the inner and outer barrier for actinide
nuclei corresponding to the local maxima along the LEP encountered at moderate
and large deformation. Experimental information is not sensitive to the 
ordering of barriers in terms of deformation, such that we discuss all results 
in terms of the primary (highest) and secondary (lowest) barriers.

\subsection{BSkG versus BSk model ingredients}
\label{sec:BSk_comparison}

All terms of the functional underlying BSkG1 and BSkG2 are essentially standard.
With the extensions discussed in the preceding subsections, the new BSkG3 model 
has now reached parity with the last entries in the BSk-model series~\cite{Goriely16} 
in terms of the formal properties of all relevant equations.
Among the BSkG-models, BSkG3 is thus the most directly comparable to 
BSk30, BSk31 and BSk32. Several differences nevertheless remain, 
primarily due to the different numerical set-up as discussed in 
detail in Ref.~\cite{Scamps21}: the three-dimensional coordinate representation 
gives the BSkG models (i) an extended reach in terms of the symmetries imposed 
on the nuclear configuration, (ii) an improved numerical accuracy but also 
(iii) a different discretisation of single-particle states in the continuum, 
and hence, pairing properties when compared to the BSk-models. Not all 
aspects of pairing are necessarily different, however: intriguingly, mass fits 
that account for pairing self-energy result naturally in low values of the 
pairing cutoff: $E_{\rm cut} = 6.5$ MeV for BSk30-31-32~\cite{Goriely16}. 
This is the natural size for coordinate space implementations such as ours, but 
this value is several times smaller than the cutoffs typically employed with 
numerical implementations that rely on an expansion in terms of harmonic 
oscillator basis states. Low values of the cutoff also arise naturally if 
one requires the pairing terms in the limit of $\rho\rightarrow 0$ to be 
compatible with the experimental neutron-neutron  $^{1}S_0$ scattering length,
see the discussion in Refs.~\cite{esbensen1997,garrido1999,chamel2010}. 

Other than those due to the numerical representation, there remain four minor 
differences that separate BSkG3 from BSk30-31-32. First is the precise 
role of the parameters $\kappa_q$: since in Ref.~\cite{Goriely16} these determined
the size of the pairing terms involving gradients in an absolute sense, while 
here they get multiplied by the deduced $V_{q}(\rho_n, \rho_p)$. The result is 
that the values of these parameters for the BSk30-31-32 models are not directly 
comparable to the ones we obtain here. 
Contrary to the older models, the form of Eq.~\eqref{eq:micro_strength} 
ensures that all pairing terms are consistently renormalized when the 
value of $E_{\rm cut}$ is changed. With this feature, our new EDF now becomes 
more suitable for time-dependent HFB calculations: only with much larger 
cutoff values can one reliably follow the dynamical evolution on the 
timescales relevant to applications, see Refs.~\cite{magierski2019,Pecak_2021}.
Aside from their relevance 
to the description of reactions of finite nuclei and nuclear fission in 
particular~\cite{Hashimoto16,Bulgac2016,Magierski2017}, such approaches are also 
of interest to explore various phenomena in NSs, for instance, the 
dynamics of neutron superfluid vortices in the crust~\cite{wlazlowski2016}. 

Second, all BSk models starting from the very first~\cite{Samyn02} 
manually enlarge the pairing strength by about 5\% for nucleon species $q=p,n$ 
when the corresponding particle number $N_q$ is odd. For BSk30-31-32, this is 
accomplished through parameters $f^{+}_{q} = 1$ and $f^{-}_q \approx 1.05$ that 
multiply $g_q(\rho_n, \rho_p)$ in the equivalent of Eq.~\eqref{eq:pairing_energy} 
in Ref.~\cite{Goriely16} when $N_q$ is even or odd, respectively. This 
recipe was formally motivated in Ref.~\cite{Nayak95} as a phenomenological 
way to (i) account for time-reversal symmetry breaking and (ii) the residual
interaction between the odd neutron and odd proton in odd-odd nuclei and 
led in practice to improved mass fits. We have opted not to employ this recipe
for several reasons. First, the factors $f_{q}^{\pm}$ cannot directly be linked to
an effective interaction. Second, we now explicitly account for time-reversal
symmetry breaking as we did with BSkG2. Third, we have recently shown that BSk31, BSkG1 and BSkG2 
all fail to describe the global trends of $\Delta^{(3)}_{np}$, specific mass 
differences that are linked to the residual $np$-interaction 
in odd-odd nuclei~\cite{Ryssens22b}. The evidence of Ref.~\cite{Nayak95} and the
BSk-models indicates that including the $f^{\pm}_q$ factors does lead to 
somewhat improved mass fits, but at the cost of two additional parameters that 
are entirely phenomenological in nature: we prefer not to include them here for the
sake of predictive power. We defer to a future study the investigation of more 
microscopic ways to reproduce at least qualitatively the  $\Delta^{(3)}_{np}$ 
mass differences and make up the difference in rms deviation of the
masses globally.

Third, our recipe to extend the INM pairing gaps to arbitrary values of asymmetry
$\delta$ is different than that employed for BSk30-31-32. During the initial 
phases of this work,  we discovered a flaw in the recipe of Ref.~\cite{Goriely16}: 
it leads to negative pairing gaps when $|\delta|$ is large but not equal to 1. 
Our new choice, Eq.~\eqref{eq:interpolation_INM}, is guaranteed to produce 
strictly positive values of the pairing gaps for both nucleon species for all
values of $\delta$. 

A final difference concerns our treatment of the mean-field potentials: as 
$g_q(\rho_n, \rho_p)$ depends on the nucleon densities, the pairing terms of the 
EDF in Eq.~\eqref{eq:pairing_energy} should contribute to the mean-field 
potentials, see for instance the appendix of Ref.~\cite{Chamel08}. We neglect 
here this coupling between the pairing and particle-hole parts of the EDF for 
simplicity, as we did for BSkG1 and BSkG2.

\section{Optimization procedure}
\label{sec:optimization}

The new model is formulated in terms of 29 parameters of which
seventeen are related to the Skyrme functional, three to the pairing 
functional ($\kappa_n$, $\kappa_p$, and $E_{\rm cut}$), five to the collective 
correction ($b$, $c$, $d$, $l$, $\beta_{\rm vib}$) and four to the Wigner energy 
($V_W$, $\lambda$, $V^{'}_W$ and $A_0$) 
\footnote{For completeness, we detail the way the Skyrme functional depends 
on these parameters in~\ref{app:couplingconstants} and provide formulas for the 
rotational and vibrational correction as well as the Wigner energy
 in~\ref{app:corrEnergies}.}.
BSkG2 relied on 25 parameters;
the new model adds six new parameters that characterize the additional terms in the 
Skyrme functional ($t_4, t_5, x_4, x_5, \beta, \gamma$) parameters and two new
pairing parameters ($\kappa_p, \kappa_n$) but no longer 
includes four parameters ($V_{\pi n},V_{\pi p},\eta, \xi$) that specified the 
more phenomenological pairing terms of the preceding model.

The primary ingredient of the objective function of our parameter adjustment 
is the set of 2457 experimental nuclear atomic masses with $Z \geq 8$ from 
AME20~\cite{AME2020}. Adjusting 29 parameters on thousands of data points
that all require at least one MOCCa calculation would be infeasible if approached
naively. Here, we rely again on the machine learning techniques developed for
the adjustment of BSkG1~\cite{Scamps21}: we train several Multi-Layer Neural 
Networks (MLNNs) to serve as emulators of MOCCa. A growing library of MOCCa
calculations serves to train the members of the committee, which then jointly
predict new parameter values of increased fitness that serve for further training. 
Once we reach an optimal set of parameter values, we perform complete calculations with MOCCa to 
obtain all the results we discuss below. We emphasize that we employ the 
MLNNs solely as a way to accelerate the parameter adjustment; the resulting model 
relies in no way on interpolation or extrapolation through machine learning.

Aside from the known masses, we also fit the average pairing gaps 
$\langle uv\Delta \rangle_q$ ($q=p,n$) to experimental values of the five-point neutron
and proton gaps $\Delta^{(5)}_{q}$ for nuclei that are sufficiently 
far from the spherical shell closures~\cite{Bender00,Scamps21}. 
We include these quantities in the objective function\footnote{
We include the pairing gaps in the objective function with a weight of 19$\%$ relative to the absolute values of the masses.} 
to control the overall
size of the pairing parameters $\kappa_n$ and $\kappa_p$, since a fit to only 
the masses tends to produce unrealistically large pairing strengths that would 
deteriorate the quality of our model for other observables~\cite{Goriely06}.
In this respect, the approach we adopt here for the pairing parameters differs from that used for BSkG2 in two ways. First,
we now control the neutron and proton pairing strengths as opposed to 
just the neutron one. Second, we do not fit the proton gaps of odd-$Z$ 
isotopic chains nor the neutron gaps of odd-$N$ isotonic chains: the $\Delta^{(5)}_q$ 
along such chains involve the binding energy of odd-odd nuclei and are for that
reason systematically smaller than the gaps in neighbouring even-$Z$ or even-$N$
chains. 
This leaves the objective function with 957 and 1153 experimental values for $\Delta^{(5)}_p$ and $\Delta^{(5)}_n$, respectively.
As discussed briefly in Sec.~\ref{sec:BSk_comparison}, current 
large-scale microscopic mass models do not account for this 
effect~\cite{Hukkanen22,Ryssens22b}. Pending a dedicated study, we prefer to 
remove the affected mass differences from the objective function in order 
to not contaminate the final parameter set. 

To include the physics of large deformation in general and fission in particular, 
we employ the two-step procedure that was originally devised for BSk14 and that 
we employed for BSkG2~\cite{Goriely07,Ryssens22,Ryssens23}. 
We repeat here only the key points: we perform an initial fit that includes
only experimental information on ground state properties. Using the optimal
parameter values resulting from this first step, we calculate fission
barriers and isomeric excitation energies as outlined in Sec.~\ref{sec:numerics}.
Freezing all other parameters, we continue fitting in a second step 
the nine parameters of the collective correction to 
reproduce at best, simultaneously, ground state properties, the RIPL-3 empirical barriers, and the known 
isomeric excitation energies. For BSkG2, we limited this fit to twelve
even-even nuclei; with the experience of Ref.~\cite{Ryssens23} under our belts, 
we now include the barriers of all 45 nuclei with $Z \geq 90$ that 
are listed in RIPL-3 as well as their 28 known isomeric excitation energies~\cite{Samyn04}.

At all stages of the parameter adjustment, we control the INM 
properties of the parameterisation in several ways. First, we fix several of the
INM properties at saturation density: we set the symmetry energy coefficient 
$J = 31$ MeV and we restrict the incompressibility modulus $K_v$ 
and the isoscalar effective mass $M^*_s/M$ to the intervals $[230,250]$ MeV \cite{Colo04}
and  $[0.8,0.86]$ \cite{Cao06b}, respectively. We also adjust $k_{\rm F}$ to reproduce
the global trend of known charge radii. The spirit of these choices is 
identical to the ones made for BSkG1 and BSkG2, but our 
value of $J$ is different.
We set $J^{\rm BSkG1/2} = 32$ MeV before by the desire to have a NeutM EoS with some
minimal degree of stiffness, though this did not suffice to render the 
previous models consistent with the existence of heavy pulsars. 
For the extended Skyrme form we employ here, this
constraint no longer applies and $J=31$ MeV was shown to be close to optimal 
for the global description of masses and particularly well-suited to reproduce 
the binding energies of neutron-rich nuclei in Ref.~\cite{Goriely16}. We confirmed
this with exploratory calculations, but did not consistently include $J$ as
a free parameter in the optimization. Our control of $K_{\nu}$ was facilitated 
by our choice of the exponents of the various density dependencies: the parameters
$\alpha, \beta$ and $\gamma $ were chosen as in Ref.~\cite{Goriely16} and 
were not actively adjusted.

A second aspect of control is new: we restrict the energy of NeutM 
at $\rho = 1$ fm$^{-3}$, i.e. at high density prevailing in NS cores, 
to the interval $[550,600]$ MeV, 
ensuring in this way a sufficiently stiff EoS and hence the compatibility of our
model with the existence of pulsars with masses above $2 M_{\odot}$.

In summary, the objective function and the parameter adjustment strategy of 
BSkG3 are similar but not identical to those of BSkG2~\cite{Ryssens22}. 
The differences are: (i) the inclusion of proton pairing gaps but removal
of neutron (proton) pairing gaps along odd-$Z$ isotopic (odd-$N$ isotonic) chains, 
(ii) a larger selection of fission data that now also encompasses odd-mass 
and odd-odd nuclei and (iii) added a constraint on the energy of NeutM 
at high density.

\section{Properties of atomic nuclei}
\label{sec:BSkG3}

\subsection{The BSkG3 parameterisation }
\label{sec:bskg3param}

\begin{table}[t]
\caption{
The BSkG3 parameter set: seventeen parameters determining the self-consistent
mean-field energy $E_{\rm HFB}$, three to the pairing 
functional, and nine determining the correction energy 
$E_{\rm corr}$. For comparison, we include the values of the BSkG2 parameter
set~\cite{Ryssens22}. Note that instead of parameter $x_2$ we list the 
values of the product $x_2 \, t_2$.
}
\begin{tabular}{l|d{6.8}|d{6.8}|d{6.8}}
\hline
    Parameters                            & {\rm BSkG2}  &  {\rm BSkG3}  \\
\hline
$t_0$         [MeV fm$^3$]               & -1885.74      & -2325.35    \\
$t_1$         [MeV fm$^5$]               &   343.59      &  749.82 \\
$t_2$         [MeV fm$^5$]               &  -8.04132     &  0.01 \\
$t_3$         [MeV fm$^{3 + 3\alpha}$]   & 12358.4       & 14083.45\\
$t_4$         [MeV fm$^{5 + 3\beta}$]    &               & -498.01 \\
$t_5$         [MeV fm$^{5 + 3\gamma}$]   &               & 266.52 \\
$x_0$                                    &     0.181775  &  0.558834\\
$x_1$                                    &    -0.584003  &  2.940880 \\
$x_2 t_2$     [MeV fm$^5$]               &  -162.003     & -432.256954 \\
$x_3$                                    &     0.101596  & 0.628699 \\
$x_4$                                    &               &  5.657990 \\
$x_5$                                    &               &  0.396933 \\
$W_0$         [MeV fm$^5$]               &   108.655     & 119.735 \\
$W_0'$        [MeV fm$^5$]               &   108.603     & 78.988 \\
$\alpha$                                 &     0.3       & 1/5 \\
$\beta$                                  &               & 1/12 \\
$\gamma$                                 &               & 1/4 \\
\hline
$V_{\pi n}$   [MeV ]               &  -483.366     &   \\
$V_{\pi p}$   [MeV ]               &  -503.790     &   \\
$\eta$                                   &     0.486     &  \\
$\xi$                                    &    0.796      &  \\
$\kappa_{n}$ [fm$^8$]                    &               & 123.20 \\
$\kappa_{p}$ [fm$^8$]                    &               & 129.07  \\
$E_{\rm cut}$ [MeV]                      &     7.998     & 7.961 \\
\hline
$b$                                      &     0.878    &  0.810 \\
$c$                                      &     8.293    &  7.756  \\
$d$                                      &     0.595    &  0.289\\
$l$                                      &     4.555    &  5.499 \\
$\beta_{\rm vib}$                        &     0.788    &  0.827 \\
\hline
$V_W$         [MeV]                      &    -1.805    & -1.716  \\
$\lambda$                                &   252.17     & 437.20  \\
$V_W'$        [MeV]                      &     0.745    & 0.502 \\
$A_0$                                    &    35.496    & 37.801 \\
\hline
\end{tabular}
\label{tab:param_skyrme}
\end{table}

\begin{table}[h]
\centering
\caption{Root-mean-square 
         ($\sigma$) and 
         mean  ($\bar \epsilon$) deviations between experiment and predictions for the BSkG2 and BSkG3 models. 
         The first block refers to the nuclear ground-state properties and the second one to fission 
         properties. More specifically, these 
         values were calculated with respect to 2457 known masses ($M$)~\cite{AME2020} 
         of nuclei with $Z$, $N \geq 8$, the subset of 299 known masses $M_{nr}$ of neutron-rich nuclei with $S_n \le 5$~MeV, 2309 neutron separation energies 
         ($S_n$), 2173 $\beta$-decay energies ($Q_\beta$), 810 measured charge radii 
         ($R_c$) \cite{Angeli13}, 45 empirical values for primary ($E_{\rm I}$)
         and secondary ($E_{\rm II}$) fission barrier heights~\cite{Capote09} 
         and 28 fission isomer excitation energies ($E_{\rm iso}$) of actinide       nuclei~\cite{Samyn04}. The first line gives the model error \cite{Moller88} on all the measured masses. } 
\begin{tabular}{l|d{6.8}|d{6.8}|d{6.8}}
\hline
\hline
Results                       &   {\rm  BSkG2}  &  {\rm  BSkG3} \\ 
\hline
$\sigma_{\rm mod}(M)$ [MeV]           &  0.668 &  0.627 \\ 
$\sigma(M)$ [MeV]                     &  0.678 &  0.631  \\ 
$\bar \epsilon (M)$ [MeV]             & +0.026 &  +0.080  \\ 
$\sigma(M_{\rm nr})$ [MeV]                &  0.851 &  0.660  \\ 
$\bar \epsilon (M_{\rm nr})$ [MeV]        & +0.308 &  -0.011  \\ 
$\sigma(S_n)$ [MeV]                   &  0.500 &  0.442 \\ 
$\bar \epsilon (S_n)$ [MeV]           & -0.006 & +0.009   \\ 
$\sigma(Q_\beta)$ [MeV]               &  0.619 & 0.534  \\ 
$\bar \epsilon (Q_\beta)$ [MeV]       & -0.019 & +0.021   \\ 
$\sigma(R_c)$ [fm]                    & 0.0265  & 0.0237  \\ 
$\bar \epsilon (R_c)$ [fm]            & +0.0007 & +0.0006  \\ 
\hline
$\sigma(E_{\rm I})    $  [MeV]        &  0.44   &   0.33    \\
$\bar{\epsilon}(E_{\rm I})  $  [MeV]        & +0.24   &  +0.06    \\
$\sigma(E_{\rm II})   $  [MeV]        &  0.47   &   0.51    \\
$\bar{\epsilon}(E_{\rm II}) $  [MeV]        & +0.10   &  +0.01    \\
$\sigma(E_{\rm iso})  $  [MeV]        &  0.49   &   0.36    \\
$\bar{\epsilon}(E_{\rm iso})$  [MeV]        & -0.36   &  -0.05    \\
\hline
\hline
\end{tabular}
\label{tab:rms}
\end{table}

Table \ref{tab:param_skyrme} contains the values of all 29 parameters 
that characterize the BSkG3 model: the first group of 17 parameters determines 
the Skyrme part of the functional, the second group consists of 3 parameters
specifying the pairing terms, the third group of 5 parameters and the final group of four parameters governs the collective correction and the Wigner
energy, respectively. For comparison, Table~\ref{tab:param_skyrme} also 
contains all parameter values of BSkG2~\cite{Ryssens22}, although four of
its pairing parameters ($V_{\pi n}$, $V_{\pi p}$, $\eta$ and $\xi$) do not have a counterpart
in the new model.

Table~\ref{tab:rms} showcases the global 
performance of the model for atomic nuclei. Concerning ground state 
properties, it lists the mean $(\bar{\epsilon})$ and rms ($\sigma$) deviations 
of the model with respect to all 2457 known masses of nuclei with $N,Z \geq 8$ 
in AME20, as well as the known neutron separation energies and $\beta$-decay 
energies from the same mass evaluation and 810 known charge radii from Ref.~\cite{Angeli13}\footnote{Note that only 810 experimental charge radii in Ref.~\cite{Angeli13} are considered here, excluding their Re, Po, Rn, Fr, Ra, Am and Cm values derived from systematics.}. We also show the mean and 
rms deviations with respect to 45, i.e. all reference values with $Z\geq 90$ from the RIPL-3 database~\cite{Capote09}
for the primary ($E_{\rm I}$) and secondary $(E_{\rm II})$ barriers. Finally, 
the table lists the deviations for the subset of 28 actinide nuclei for which 
the excitation energy of a fission isomer ($E_{\rm iso}$) is known.

BSkG3 presents a modestly (though not negligible) improved description over BSkG2 for ground-state
properties: the rms deviation for the known masses and charge radii is, respectively, 6\% and 12\% smaller than its predecessor. The global improvements for 
the $S_n$ and $Q_{\beta}$ values amount to roughly 12\% and 15\%, respectively. This is
not a trivial observation, since an improved description of the absolute 
values of the masses does not guarantee a reduced rms deviation for mass differences.
Since the BSkG3 model ingredients differ in several ways from the BSkG2 ones,
it is not possible to unambiguously pinpoint a single source for this improvement. 
Nevertheless, we note that both the microscopic treatment of pairing~\cite{Goriely09}, 
and the reduced value of $J$~\cite{Goriely16} have shown to lead to a reduction 
of the rms deviation on the masses in the past.


The  description of known absolute binding energies that BSkG3 offers does not quite 
reach the level of the later entries in the BSk-series; in particular BSk27
achieves an rms deviation of 0.512 MeV with respect to the AME20 
masses~\cite{Goriely13b}. BSk27 is based on a standard Skyrme form and, like 
BSkG1 and BSkG2, is not compatible with the existence of heavy pulsars. Later 
models such as BSk31 ($\sigma(M) = 0.585$ MeV) 
are less accurate but satisfy this constraint. BSkG3 and BSk31 share (almost) all 
model ingredients such that a comparison between both parameterisations is more direct.
The difference in rms deviation between BSkG3 and BSk31 is but a few tens of 
keV, i.e. roughly 8\%, but still significant. Without additional costly parameter fits, 
we cannot pinpoint the exact origin of this difference but we surmise it is mostly due to
our omission of the phenomenological pairing factors $f^{\pm}_{p/n}$ employed
by the BSk-series as discussed in Sec.~\ref{sec:BSk_comparison}. 
However, when concentrating on the subset of 299 masses of neutron-rich nuclei with 
$S_n\le 5$~MeV (of particular interest for the extrapolation towards the exotic neutron-rich 
nuclei involved in the $r$-process nucleosynthesis), the superiority of the BSkG3 model with respect to BSkG2 is striking, with a reduction of about 200~keV. The BSkG3 predictions for this subset also appear to be 
more accurate than those obtained with BSk27 ($\sigma(M_{nr})=0.685$~MeV) and BSk31 ($\sigma(M_{nr})=0.733$~MeV).
When dealing with mass differences, BSkG3 describes the $S_n$ and $Q_{\beta}$ values with accuracy rather similar to (or even better than) the one found by the latest BSk models, in particular $\sigma(S_{n})=0.429$~MeV and 0.479~MeV and $\sigma(Q_{\beta})=0.525$~MeV and 0.597~MeV, for BSk27 and BSk31, respectively.

We argued in Ref.~\cite{Ryssens23} that BSkG2 offered the best all-round 
global description of static fission properties: its simultaneous description of
primary and secondary barriers, as well as isomeric excitation energies of 
the 45 $ Z\geq 90 $ actinide nuclei with an rms deviation of less than 0.500~MeV 
was unmatched in the literature at the time of writing.
However, the new BSkG3 model does significantly better in several ways. First, 
the primary fission barriers, which remain the most important quantity from the 
point of view of applications, are reproduced with an rms deviation of only
0.330~MeV, i.e. a reduction of 25\% compared to BSkG2. Only the PC-PK1~\cite{Lu14}
model outperforms BSkG2 in describing primary barriers: it achieves an rms deviation
of 0.37 MeV but results are only available for twelve even-even nuclei. The rms 
deviation of BSkG3 for primary barriers is, to the best of our knowledge, the 
lowest ever reported in the literature. 
Also, the rms deviation for fission isomers has decreased by a similar
amount. Second, the mean deviations $\bar{\epsilon}$ have been reduced dramatically 
for all three fission properties: a factor of two for the secondary barriers, 
four for the primary barriers and even nine for the isomer excitation energies.
The only (minor) downside is a
slightly larger rms deviation for the secondary barriers: $\sigma(E_{\rm II}) = 0.51$ MeV.  
We will discuss the BSkG3 fission properties and the reasons for this improved
performance in more detail in Sec.~\ref{sec:fission_discussion}, but mention 
already here that it is not due to the incorporation of additional parameters
related to fission.

We end this discussion of global performance with a remark on the neutron skin of $^{208}$Pb, 
 which the new model predicts to be 0.16 fm, a value somewhat smaller than the 
 value obtained with BSkG1 and BSkG2 (0.18 fm). This is as expected from our choice to implement a lower symmetry energy coefficient $J$ compared to the 
 preceding models. 
 Both values remain in good 
 agreement with constraints deduced from measurements of the electric dipole 
 polarisability~\cite{rocamaza15}, the antiproton capture of $^{208}$Pb~\cite{klos07}, and ab-initio predictions~\cite{Hu22}. 
The authors of Ref.~\cite{Giacalone23} have recently derived the  $^{208}$Pb neutron skin from ultrarelativistic heavy collision of  $^{208}$Pb + $^{208}$Pb. They obtain $\Delta r_{\rm np} = 0.217 \pm 0.058$ fm, our prediction is therefore marginally consistent within the lower limit of this result.  
Nonetheless, BSkG1-3 predictions differ from the value of $0.283 \pm 0.071$~fm reported by the PREX-II collaboration \cite{PREX21} extracted from
  measurements of the parity-violating asymmetry in the elastic scattering of 
  longitudinally polarized electrons on $^{208}$Pb. The PREX-II value 
  remains noticeably larger than those derived from the above-mentioned experiments and from theoretical predictions.

\subsection{Nuclear masses}

\begin{figure}
  \includegraphics[width=0.45\textwidth]{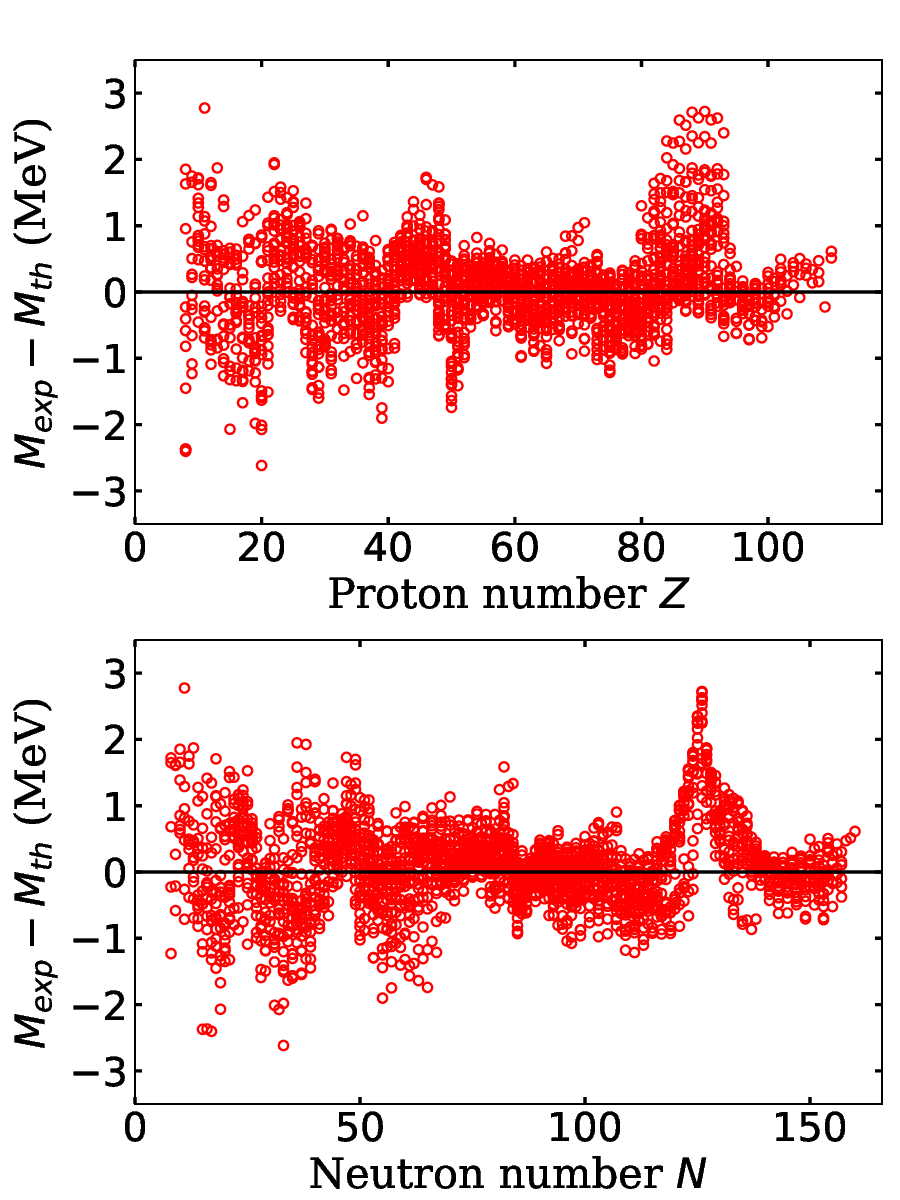}
\caption{Differences between experimental \cite{AME2020} and BSkG3 masses as a function of the proton number (top) and neutron number (bottom).}
\label{fig:massexp}   
\end{figure}

We show the difference between experimental and calculated
masses in Fig.~\ref{fig:massexp} as a function of proton number (top panel)
and neutron number (bottom panel). 
The BSkG3 model achieves a general good fit
to the data with only a few nuclei exhibiting a deviation that is larger than 2 MeV. The largest deviations concern either light nuclei or nuclei close 
to the magic numbers; these patterns are similar to those of BSkG- and 
BSk-families of models~\cite{Goriely16,Ryssens22}.

%

\begin{figure}
  \includegraphics[width=0.45\textwidth]{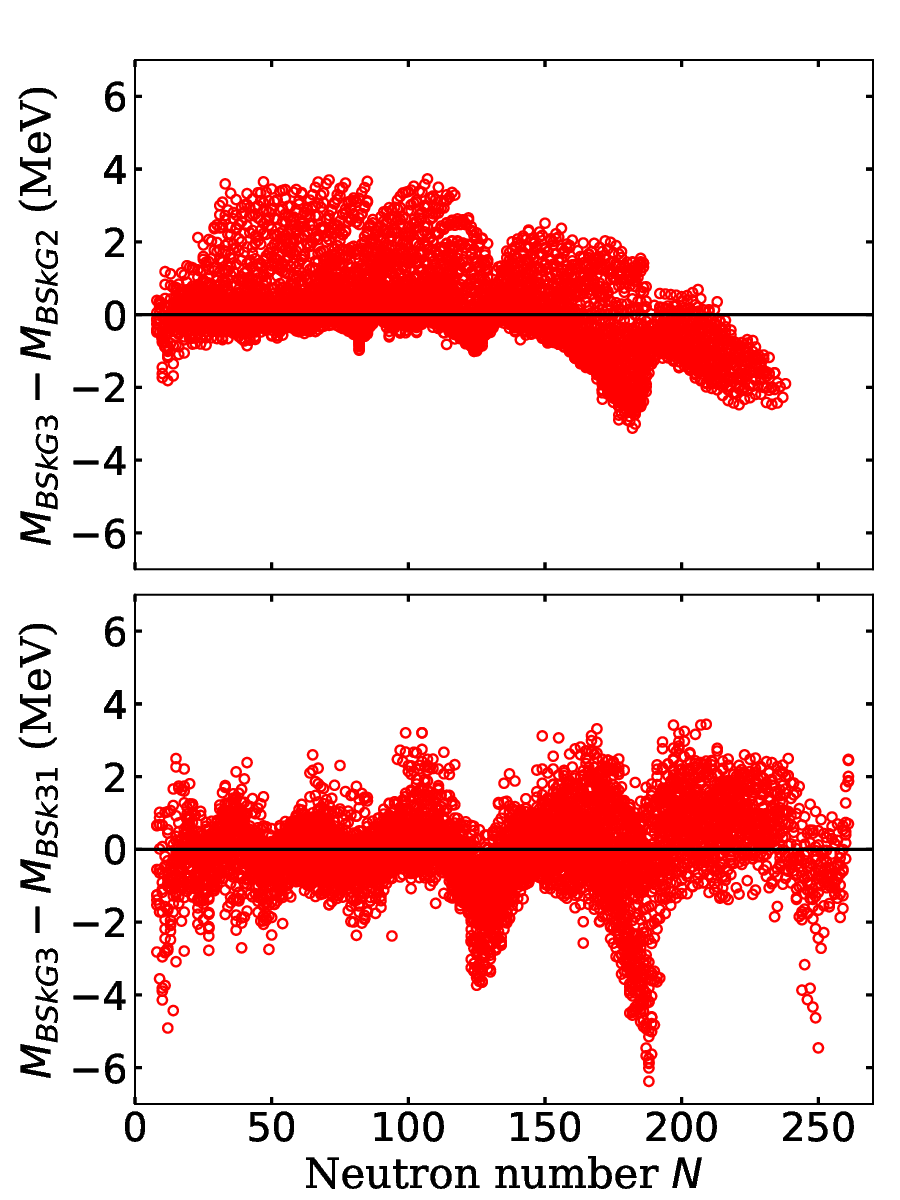}
\caption{Mass differences between BSkG3 and BSkG2\cite{Ryssens22}(top) and BSkG3 and BSk31~\cite{Goriely16}(bottom) as a function of the neutron number $N$ for all nuclei with $8\le Z \le 118$ lying between the BSkG3 proton and neutron drip lines.}
\label{fig:massBskgs}   
\end{figure}

In Fig.~\ref{fig:massBskgs} the difference between the masses obtained with 
BSkG3 and BSkG2 (top) or BSk31
(bottom) are displayed as a function of 
the neutron number for all nuclei with $8 \leq Z \leq 118$ lying between the BSkG3 proton and neutron drip lines. 
We note that BSkG3 produces binding energies that are, on average, 
smaller than the ones obtained from BSkG2 for intermediate and light nuclei.
For heavy nuclei BSkG3 obtains binding energies predominantly larger than BSkG2. These differences are rather smooth, due to the similar predictions for shell gaps among BSkG2 and BSkG3, as noted in Fig.~\ref{fig:shellgaps}.
These differences, however, never exceed 4 MeV.
When compared to BSk31 we note a larger difference between models, especially close to the $N=126$ and $N=184$ shell closures where it can reach up to 6 MeV. 
This can be understood from the stronger shell effects for this model (see Fig.~\ref{fig:shellgaps}) which is a consequence of the lower isoscalar effective mass of BSk31, as seen in Tab.~\ref{tab:nucmatter}.


%

\subsection{Shell structure}

\begin{figure*}
\centering
  \includegraphics[width=0.75 \textwidth]{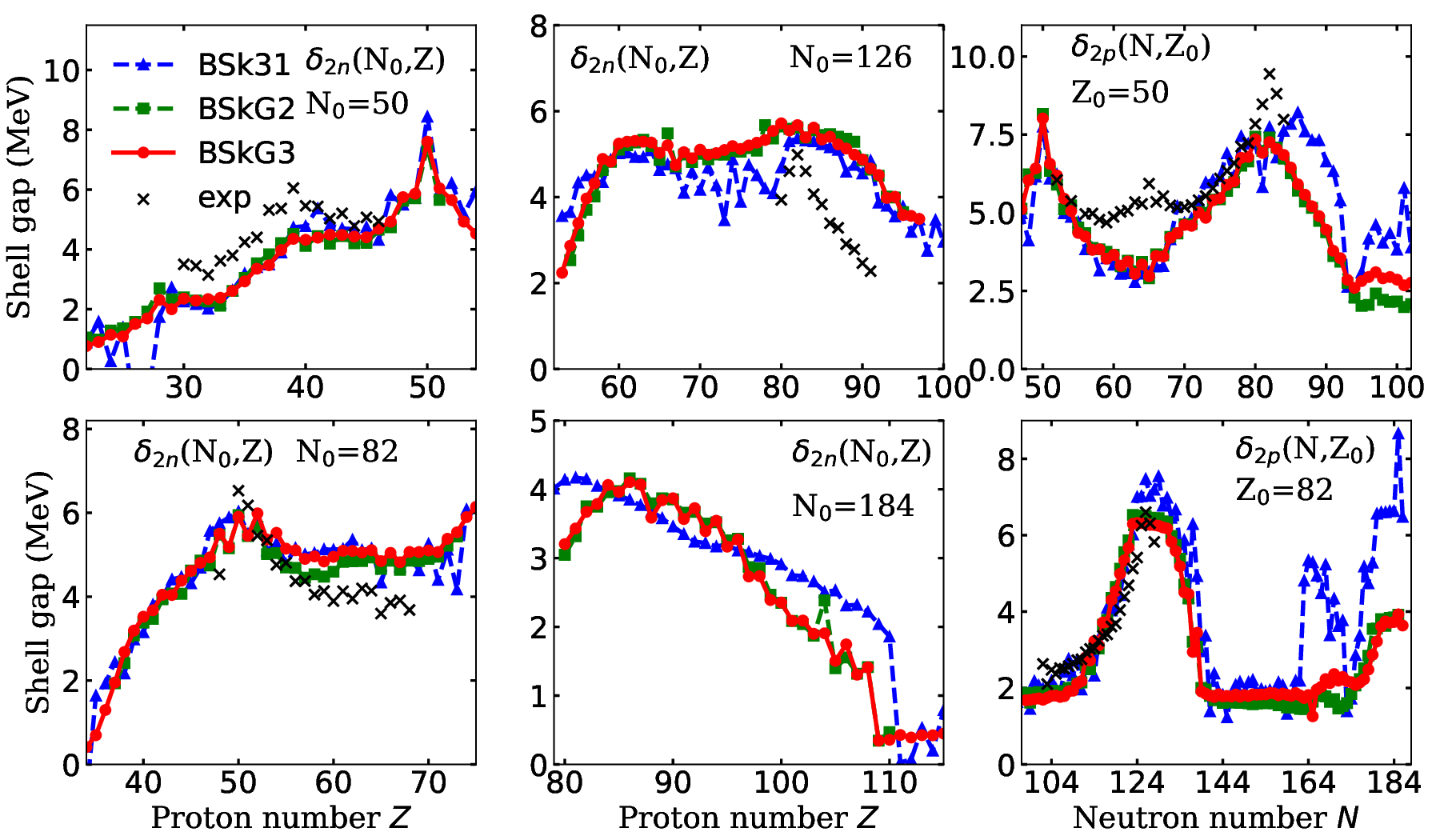}
\caption{Shell gaps predicted by BSk31 (blue triangles), BSkG2 (green squares) 
       and BSkG3 (red circles) for N=50, N=82, N=126, N=184, Z=50 and Z=82. Results are 
       compared with available data from AME20~\cite{AME2020} (black crosses).
     }
    
\label{fig:shellgaps}       
\end{figure*}

The shell effects can be investigated by the neutron and proton shell gaps defined as,
\begin{eqnarray}
\delta_{2n} (N_0 , Z ) &=&  S_{2n} (N_0 , Z ) - S_{2n} (N_0 + 2, Z ) , \label{eq:shellsn}\\
\delta_{2p} (N , Z_0 ) &=&  S_{2p} (N , Z_0 ) - S_{2p} (N, Z_0 + 2 ) , 
\label{eq:shellsp}
\end{eqnarray}
where $ S_{2n/ p}$ are the two-neutron/-proton separation ener\-gies. The $ \delta_{2n/ p}$ are indicators for shell
closures, but they are not measures of the size of the
gap in the single-particle spectrum as they are sensitive
to any structural change between the three nuclei entering Eqs.~(\ref{eq:shellsn}-\ref{eq:shellsp}) \cite{Bender08}. 

We show in Fig. \ref{fig:shellgaps} the $ \delta_{2n/ p}$ values across 
spherical shell closures for the BSkG2, BSkG3 and BSk31 models for the
chains of heavy semi-magic nuclei, as well as available experimental data. 
All three models produce similar neutron shell gaps for the $N_0 = 50$ and 82 gaps, agreeing with experimental values about equally well.
We note that for $N_0 = 126$ (upper middle), BSk31 predicts smaller shell gaps for $68 < Z < 82$ with stronger odd-even staggering in comparison to BSkG2 and BSkG3.
The bottom middle panel shows the neutron gap near $N_0 = 184$ where, for the neutron rich isotopes, BSk31 produces a gap roughly $13 \%$ higher than BSkG2 and BSkG3, while the three models agree with each other for $Z > 82$.  
For $Z_0 = 50$ (upper right), BSk31 predicts stronger proton gaps for $N > 82$, reaching $\approx 25 \%$ around $N = 88$. For the $Z_0 = 82$ proton shell gap (bottom right), BSk31 produces a $\approx 15 \%$ higher gap around $N = 126$, and even a stronger gap for the most neutron rich isotopes around  $N \ge 166$. 
The difference between the BSk31 and BSkG3 shell gaps explains the large mass differences at $N \approx 126$ and $N \approx 184 $ in Fig.~\ref{fig:massBskgs}.
Despite a relatively good overall agreement between experimental and BSkG3 shell gaps, the $N_0 = 126$ neutron shell gaps remain overestimated leading to an overbinding of masses along the $^{208}$Pb isotonic chain, as clearly observed in Fig.~\ref{fig:massBskgs}. This shell gap was better described by some of the BSk models, in particular BSk27~\cite{Goriely13b}.

\subsection{Pairing properties}
\label{sec:pairing_properties}

\begin{figure}
  \includegraphics[width=0.5\textwidth]{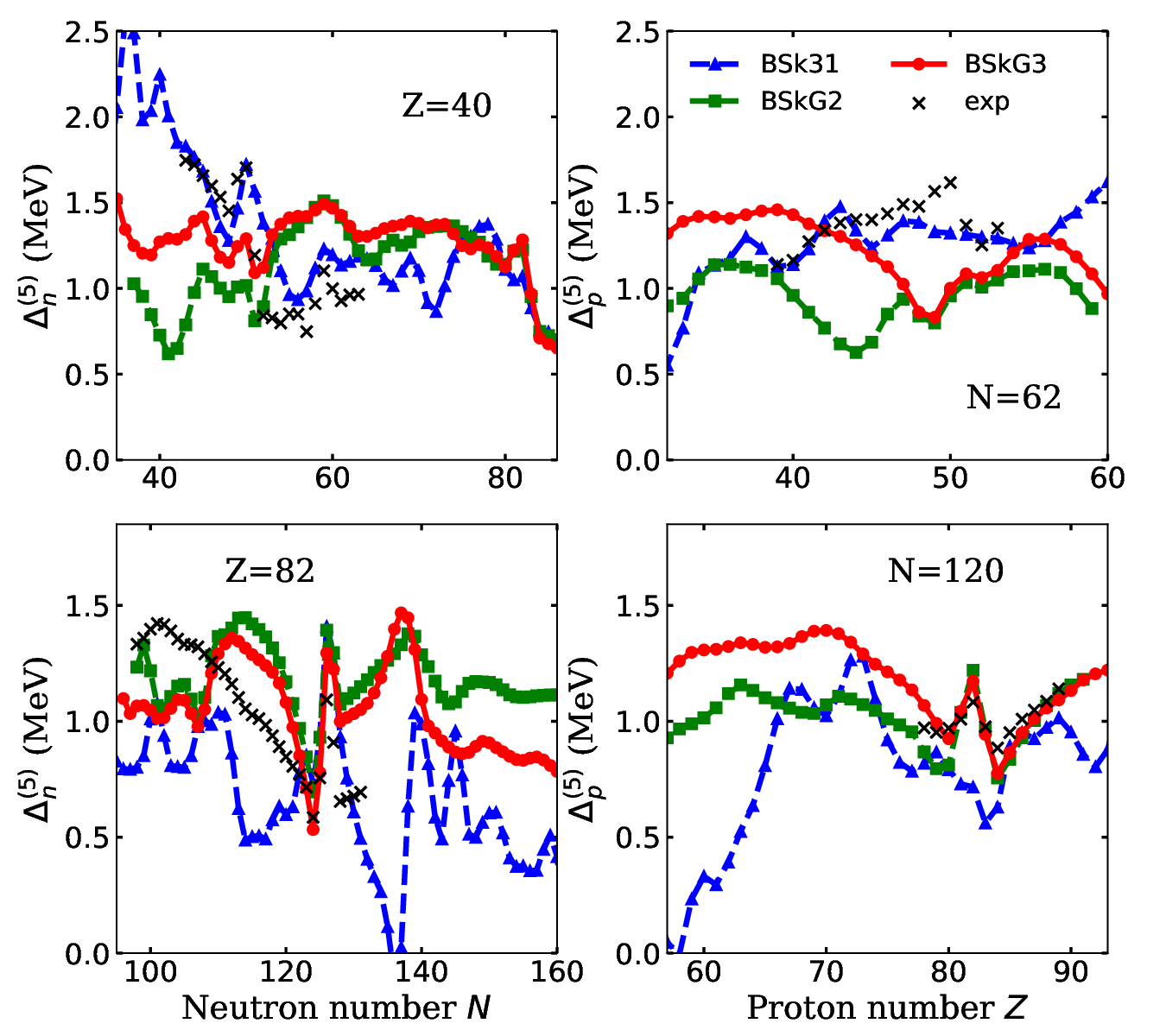}
\caption{Left: Neutron five-point pairing gaps $\Delta^{(5)}_n$ along the Sn(top) and Pb(bottom) isotopic chains as calculated with BSkG3(red circles), BSkG2(green squares) and BSk31(blue triangles). Right: Proton five-point pairing gaps $\Delta^{(5)}_p$ along the N=82(top) and N=120(bottom) isotonic chains. Black crosses show the experimental data from AME20~\cite{AME2020}. }
\label{fig:delta5}       
\end{figure}

The right (left) panels of Fig.~\ref{fig:delta5} show the five-point neutron (proton)
gaps $\Delta^{(5)}_{n/p}$ for BSkG3, BSkG2, BSk31 and available data.
The top left panel shows $\Delta^{(5)}_n$ for Zr ($Z=40$), bottom left for Pb 
($Z=82$) while right panels shows the $\Delta^{(5)}_p$ for the isotones $N=62$ (top) and $N=120$ (bottom). 
This quantity is generally indicative of the strength of 
pairing correlations in nuclei, but other effects such as time-odd terms (see discussion in Ref.~\cite{Ryssens22b}) 
and structural changes along isotopic chains also contribute~\cite{Bender00b}. 
The models predict overall similar $\Delta^{(5)}_{n/p}$ and reproduce rather well 
the experimental values. A few differences are however present. 
On top left ($Z=40$) we note that BSkG2 and BSkG3 underestimate the neutron 
pairing gaps for neutron-deficient nuclei, and BSk31 reproduces better the experimental data in this particular region. 
For $Z=82$ all models present similar predictions for the neutron-deficient region and close to the $N=126$ shell closure. For the neutron-rich region, beyond $N=126$, BSk31 presents lower values for the neutron pairing gaps when compared to BSkG's, in particular in the region $N \approx 138$ where  $\Delta^{(5)}_n$ is close to zero. 
Finally, for $N=120$ all models reproduce rather well the experimental data, they however present considerable differences in the proton-deficient region where BSk31 predicts significantly lower values for the proton pairing gaps compared to BSkG2 and BSkG3. This is directly related to the interpolation recipe used to compute pairing gaps from INM at arbitrary asymmetry values: BSk31 systematically predicts a collapse of the $\Delta^{(5)}_{p}$ for proton-deficient nuclei. Our new prescription (see Eq.~\eqref{eq:interpolation_INM} and Sec.~\ref{sec:BSk_comparison}) avoids such a collapse for extreme $N$/$Z$ values.
%

\begin{figure}
  \includegraphics[width=0.5\textwidth]{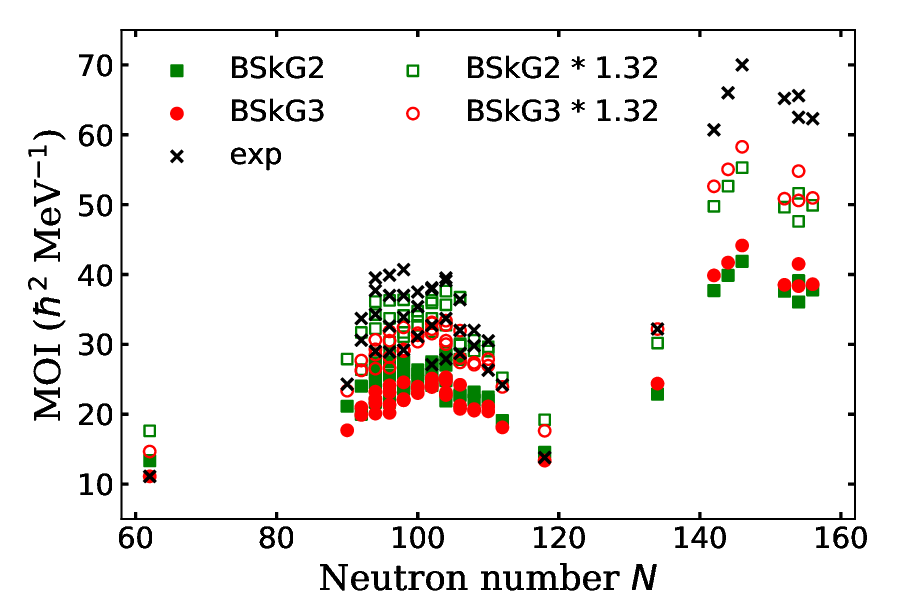}
\caption{Belyaev moments of inertia (MOI) as a function of neutron number for BSkG2 (green squares) and BSkG3 (red circles). Experimental data~\cite{Zeng94,Afanasjev00,Pearson91} is given in black squares. See text for details.}
\label{fig:MOI}       
\end{figure}

The rotational moment of inertia is another way to gauge the pairing 
properties of the new model. We compare in Fig.~\ref{fig:MOI} the 
calculated Belyaev moments of inertia (MOI) to experimental data for 
48 even-even nuclei~\cite{Zeng94,Afanasjev00,Pearson91}. Although the 
Belyaev MOI is a crucial ingredient in the collective correction, 
Eq.~\eqref{eq:Ecorr}, it remains a perturbative quantity that does not 
capture all aspects of the nuclear response to rotation. A more advanced 
calculation results in values roughly 32\% larger~\cite{Ryssens22}: we show for 
simplicity all calculated values multiplied with 1.32. With this simple 
correction taken into account, both BSkG2 and BSkG3 reproduce fairly well 
the experimental values for medium-heavy nuclei, indicating that the pairing 
properties of BSkG3 were also well-controlled in this respect. We observe though
that BSkG2 produces somewhat larger MOI values in this region. All BSkG models, 
including BSkG1 (not shown) underestimate systematically the MOI of actinide
nuclei, although it is unlikely that pairing properties are the culprit.


%

\subsection{Ground state deformations}

Our three-dimensional numerical representation naturally allows the nuclear
ground state densities to adopt a wide range of deformed, i.e. non-spherically 
symmetric, shapes that can be characterized by the multipole moments 
$\beta_{\ell m}$. In practice, we find five qualitatively different types of 
shapes: 
 (i) rotationally symmetric spheres ($\beta_{\ell m} = 0$ for all $\ell,m$)
 (ii) axially symmetric ellipsoids ($\beta_{2} > 0, \beta_{3} =0$) that are
 either prolate ($\gamma = 0^{\circ}$) or oblate ($\gamma = 60^{\circ}$),
 (iii) triaxial ellipsoids without any rotational symmetry 
       ($\beta_{2} > 0, \beta_{3} =0, \gamma \in ]0^{\circ}, 60^{\circ}[$ ) 
 (iv) reflection-asymmetric but (nearly) axially symmetric pear-like shapes 
 ($\beta_2 > 0, \beta_3 > 0, \gamma \approx 0^{\circ}$)
\footnote{Although, fruit-wise, these shapes have been said to resemble mangoes more.}
and (v) a very limited number of more general shapes that combine
triaxial deformation and reflection asymmetry. 
Although this categorisation gives an intuitive picture of the shapes 
encountered in our calculations, it is imprecise and incomplete: we repeat our
remark of Sec.~\ref{sec:numerics} that all multipole moments that are not
constrained by imposed symmetries are naturally included in the (semi-)variational
optimization of the energy. Although less important than quadrupole deformation, 
our calculations predict non-vanishing values for both $\beta_{4}$ and $\beta_{6}$
for many nuclei~\cite{Scamps21}. The multipole moment $\beta_{32}$, although 
allowed in our calculations, is seemingly not meaningfully exploited by any nucleus;
we comment on this remarkable absence below. 

\begin{figure}
 \includegraphics[width=0.5\textwidth]{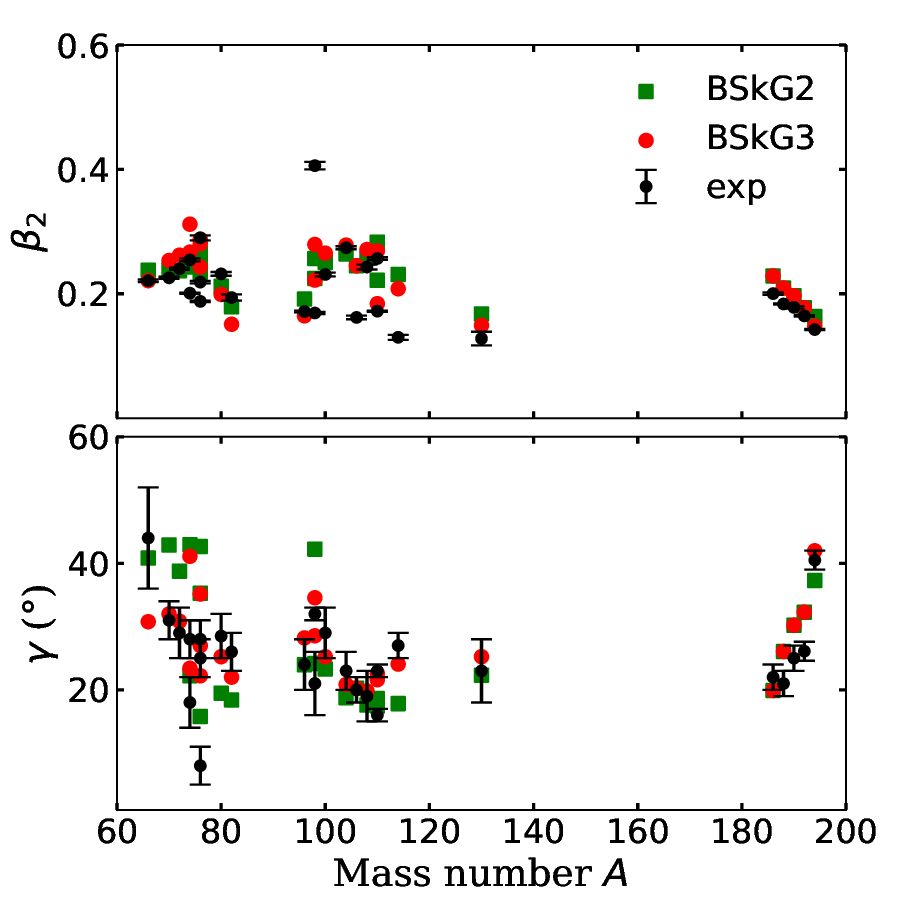}
\caption{Calculated quadrupole deformation $\beta_2$ (top panel) 
          and triaxiality angle $\gamma$ (bottom panel) with BSkG2 (green squares)
          and BSkG3 (red circles), compared to experimental information 
          (black circles with error bars).
          Top panel: quadrupole deformation $\beta_2$ with experimental
                        information from Nudat~\cite{nudat}.
                        Bottom panel: triaxiality angle $\gamma$ with 
                        experimental data points extracted from measured sets of 
                        transitional and diagonal $E2$ matrix 
                        elements~~\cite{Magda_priv,Rocchini21,Sugawara03,Ayange16,Toh00,Ayange19,Clement07,Kavka95,Clement16,ZielinskaPhD,Zielinska02,Wrzosek12,Srebrny06,Wrzosek20,Fahlander88,Svensson95,Morrison20,Wu96}.
                       }
\label{fig:deform}  
\end{figure}

We already discussed in detail the systematics of quadrupole 
deformation, both axial and triaxial, in Ref.~\cite{Scamps21} for BSkG1;
globally speaking all three BSkG-models provide comparable predictions for
this quantity. Local differences
arise for many nuclei, but a detailed study is outside the scope
of this work. Here, we only confirm that the new models succeed in
describing the (limited) available experimental information on ground 
state triaxial deformation. To this end, Fig.~\ref{fig:deform} compares 
the calculated values of $\beta_2$ (top panel) and $\gamma$ (bottom panel) 
to experimental information for 26 even-even nuclei: we compare to values in 
the NuDat database~\cite{nudat} for the total quadrupole deformation and to 
mean values of the triaxiality angle obtained from Coulomb excitation experiments~\cite{Magda_priv,Rocchini21,Sugawara03,Ayange16,Toh00,Ayange19,Clement07,Kavka95,Clement16,ZielinskaPhD,Zielinska02,Wrzosek12,Srebrny06,Wrzosek20,Fahlander88,Svensson95,Morrison20,Wu96}.
Compared to BSkG2, BSkG3 yields nearly identical results for the total 
quadrupole deformation across this range of mass number $A$ and yet provides 
a moderately improved description of the triaxiality angle $\gamma$ for 
the lightest nuclei with the exception of $^{66}$Zn, the lightest isotope
in the list.

%
%

\begin{figure*}
  \includegraphics[width=1.0\textwidth]{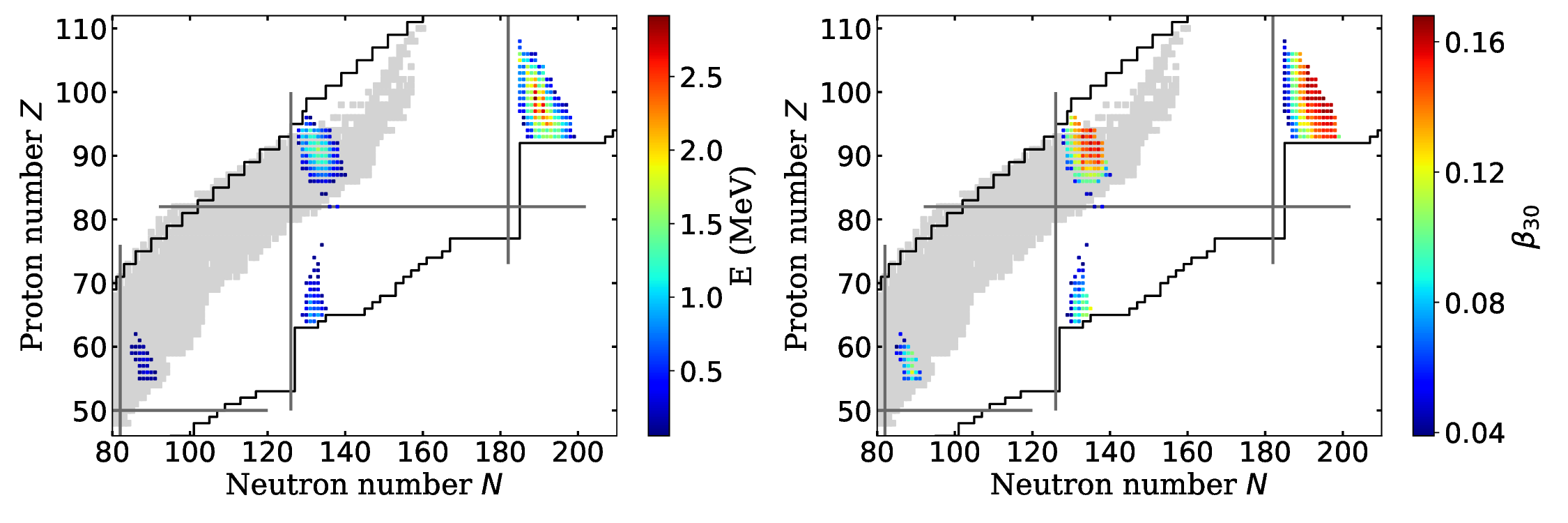}  
\caption{Left: Binding energy differences between calculations restricted to quadrupole deformations and to calculations that allow for octupole shapes. Right: octupole deformation $\beta_{30}$. Gray shadow shows the experimental nuclei and black lines indicate BSkG3 drip lines. }
\label{fig:oct}  
\end{figure*}

The possibility of reflection-asymmetric ground states is new to BSkG3: we 
show its global impact in Fig.~\ref{fig:oct}: its right panel indicates where
finite values of $\beta_{30}$ occur while its left panel show the gain in binding
energy due to reflection asymmetry, i.e. the difference in total energy between
a reflection asymmetric and a reflection symmetric calculation. Globally speaking,
the influence of reflection asymmetry is minor: we found only 196 nuclei that
gain more than 500 keV of binding energy due to the inclusion of this degree 
of freedom, the majority of which are exotic neutron-rich isotopes far beyond 
(current) experimental reach. 
The rms deviation on the masses is thus mostly 
unaffected by the inclusion of octupole deformation: an artificial BSkG3 mass
table restricted to reflection symmetric ground states produces 
$\sigma_{\rm refl. symm.}(M) = 0.620$ MeV, only 11 keV lower 
than our complete calculation and thus justifying a posteriori our choice to 
not include this degree of freedom in the parameter adjustment.

Octupole deformation becomes energetically favourable when the nucleus has 
the freedom to mix the single-particle states of a unique-parity intruder shell 
of angular momentum $j$ with those of a normal-parity $j-3$ shell~\cite{Chen21}.
This explains the distribution of octupole deformation in Fig.~\ref{fig:oct}: 
such single-particle configurations typically occur just above closed shells 
for particle numbers close to the so-called octupole magic numbers. Three of 
these, $N_{\rm oct} = 56, 88, 134$, correspond well to three regions of
octupole deformation in Fig.~\ref{fig:oct}. We find no stable octupole deformed
minima near the lightest traditional octupole magic number, $N_{\rm oct} = 34$,
likely because in such light nuclei octupole correlations are not strong
enough to produce static octupole deformation in a mean-field calculation. 
Reflection asymmetry has a strong effect on the mass of exotic neutron-rich 
nuclei some of which gain almost 3 MeV of additional binding energy. The largest
effect in our calculations is produced at $N=190$,   
which is not typically cited as an octupole magic number but is where BSkG3 
predicts that the intruder states emanating from the $k_{17/2}$ orbital mix
with the many negative parity single-particle states just above $N=184$.
Our approach to obtaining the reflection asymmetric nuclei is based on the minimum energy of both configurations. Therefore there is no guarantee of any continuous behavior in $\beta_{30}$, only in energy.
This can be noted in Fig.~\ref{fig:oct}, where $\beta_{30}$ increases at increasing $N$ and abruptly goes to zero, which is not the case for the energy differences. This effect can be understood with the following: with increasing $N$, the octupole deformed minimum and the (typically prolate) reflection symmetric saddle point move away from each other in deformation space. 
When far enough from each other, their shell structure becomes sufficiently different such that both points on the PES do not depend in the same way on $N$. At some given neutron number, the reflection symmetric point achieves a lower value for the energy than the octupole one, resulting in a smooth change of energy difference, though not of octupole deformation. 

The results shown in Fig.~\ref{fig:oct} are comparable to earlier global 
 surveys of reflection asymmetry with EDF-based models, although these have 
 all been limited to even-even nuclei while we include odd-mass and odd-odd 
 isotopes. All models that we are aware of predict the two regions of octupole
 deformation within the experimentally accessible part of the nuclear chart,
 independently of whether they are based on a Skyrme-type, Gogny-type or
 relativistic EDF~\cite{Robledo11,Agbemava2017,cao2020}. 
 When calculations
 for exotic neutron-rich systems are available, all models generally also 
 agree on the existence of the island of octupole deformation beyond 
 $N\sim 184$~\cite{Agbemava2017,cao2020}. The precise extent of the four regions
 in nucleon number varies strongly from one model to another, since whether
 or not octupole deformation appears in a given nucleus depends on  
 single-particle properties as described above, but also on the models bulk 
 properties such as its effective mass and its surface energy 
 coefficient~\cite{Ryssens2019b,Cao20,Chen21}. A few global studies of 
 reflection asymmetry in the context of macroscopic-microscopic approaches 
 are also available, but only find strong effects in the $Z \sim 88, N\sim 134$ 
 region that are nevertheless smaller than those obtained in EDF-based models.
 Although not trivial, this last observation seems natural since we established 
 that the effect of triaxiality turns out larger in self-consistent EDF-based models
 than in microscopic-macroscopic approaches~\cite{Scamps21,Ryssens23}.

The impact of reflection asymmetry on the nuclear binding energy is not
directly experimentally observable, and the presence of (static) octupole 
deformation is mostly inferred through other observables such as the appearance
of rotational bands of alternating parity and enhanced $E1$ transitions, 
or even indirectly through the trends of charge radii~\cite{Verstraelen19}
or the momentum distribution of particles emitted in heavy-ion collisions~\cite{zhang2022}. 
Most available experimental information concerns the $Z\sim 88, N \sim 134$ 
region and (to a lesser extent) the $Z\sim 56, N\sim 88$ region~\cite{Butler96}, 
in agreement with our calculations. Even though such studies on octupole 
deformation are even less numerous than those dealing with triaxiality, Coulomb 
excitation experiments such as those of Ref.~\cite{Gaffney13,Bucher2016} directly
confirm the presence of static octupole deformation in both regions.

We defer a more detailed study of reflection asymmetry to future work and limit
ourselves for this study to the effect of reflection asymmetry on binding 
energies from an astrophysical point of view. We expect that the region 
of octupole deformation near $N \sim 190 $ will be the most consequential
since the effect on the binding energy is large and typical r-process 
trajectories produce a significant population of nuclei in this region before
neutron irradiation ends. Our calculations point to two potentially significant
effects: enhanced stability with respect of fission and the modification 
of the neutron dripline in this region. All else being equal, an additional 
contribution to the binding energy of 2 MeV of a nucleus effectively enlarges 
its fission barrier by the same amount and makes the spontaneous or induced fission 
of such a system dramatically less likely. This is likely to impact r-process
simulations, since the fission properties of nuclei in this region determine 
for example the production of stable isotopes for $110 \leq A \leq 170$, but 
also impact the heating rate of kilonovae at late times through the (possible)
production of superheavy elements~\cite{Goriely15,Kullmann2022}. The effect on
the neutron dripline then simply enlarges the possibilities for the r-process
to create exotic isotopes: for the Np($Z=93$) and Pu($Z=94$) isotopes, the 
additional binding energy due to reflection asymmetry extends the dripline 
from $N=189$ to $N=209$ and $N=207$, respectively.
\footnote{There is a similar but less dramatic effect for $Z=64$ and 65: 
reflection asymmetry is responsible for extending the neutron dripline by
four neutrons.
}

To end this subsection, we remark again on the absence of any $\beta_{32}$ 
   deformation. Nuclei in our calculations are free to take non-axially symmetric
   shapes that are also asymmetric under reflection, and hence could a priori take
   advantage of this type of deformation to lower their total energy. Only a 
   handful of nuclei exploit this possibility but the associated values of $\beta_{32}$
   are tiny: the largest value in absolute sense we obtain is $-0.038$ 
   for $^{196}$Eu. We take this as an indication that this type of deformation
   is not very relevant to nuclear ground states from a global point 
   of view, although we mention that there are some indications that 
   a very limited number of isotopes could exhibit non-axial octupole deformation 
   in their ground state without any accompanying quadrupole deformation~\cite{Dudek18}.
   We refrain however from drawing too strong conclusions, since our numerical 
   search for the nuclear ground state was not completely general and possibly 
   biased by our inclusion of a temporary constraint on $\beta_{30}$ in every 
   calculation, as explained in Sec.~\ref{sec:numerics}.

\subsection{Actinide fission barriers}
\label{sec:fission_discussion}

\begin{figure}
  \includegraphics[width=0.5\textwidth]{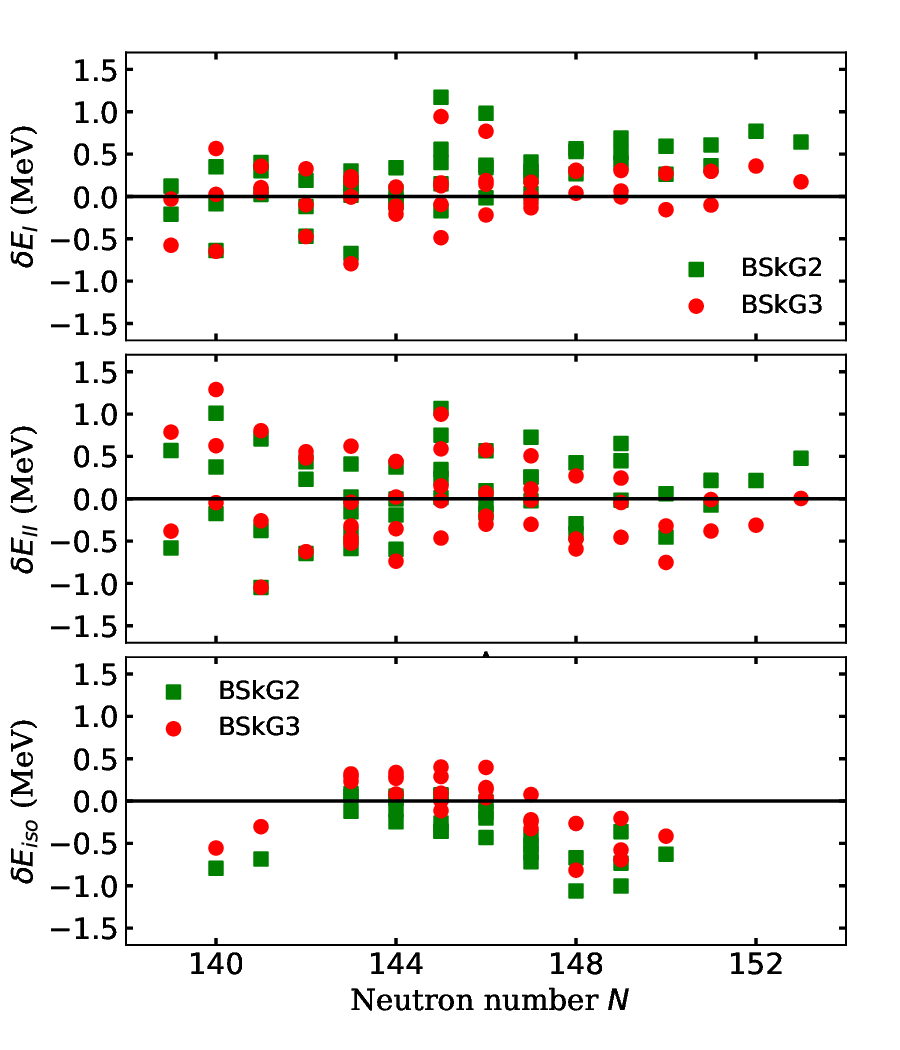}
\caption{Difference between calculated and reference values~\cite{Samyn04,Capote09} for the 
         three different static fission properties ($E_{\rm I}, E_{\rm II}$
         and $E_{\rm iso}$) as a function of neutron number for BSkG2 
          (green squares) and BSkG3 (red circles).
          }
\label{fig:fissionbar}   
\end{figure}

To study the performance of the new model for static fission properties, we show
in Fig.~\ref{fig:fissionbar} the deviations of BSkG2 and BSkG3 calculations with 
respect to the reference values: 45 primary barriers (top panel) and 45 secondary 
barriers (middle panel) of the RIPL-3 database~\cite{Capote09} and 28 isomer 
excitation energies (bottom panel). As discussed in Sec.~\ref{sec:BSkG3}, 
BSkG3 offers significantly reduced mean deviations for all three fission properties
compared to its predecessor, particularly for the isomer excitation energies. 
The new model also improves on the rms deviations for the primary barriers
and isomer excitation energies resulting in a smaller spread of the values 
in the top and bottom panels of Fig.~\ref{fig:fissionbar}, at the cost of a 
slightly less accurate description of the secondary barriers in the middle 
panel. 

The new model now describes all primary barriers for $Z \geq 90$ 
nuclei in RIPL-3 within 1 MeV. We repeat the observations of Ref.~\cite{Ryssens23b}: 
this performance with respect to the static fission properties is not universal 
among EDF-based models in the literature. When the physics of large deformation
is not included in the parameter adjustment, the predictions of such a model 
can be off by up to 10 MeV~\cite{Jodon16}. We ascribe the success of both BSkG2 
and BSkG3 to (i) the inclusion of fission properties in the parameter adjustment
and (ii) our three-dimensional calculations of the fission PES that allow for
nuclear shapes with triaxial and octupole deformation~\cite{Ryssens23}. 
The BSkG2 and BSkG3 PESes for actinide nuclei and the fission paths obtained 
from them are qualitatively similar, we do not show any of them here but refer
the interested reader to Ref.~\cite{Ryssens23}.  We only mention that, as for 
BSkG2, the fission paths of essentially all nuclei 
we consider exploit triaxial deformation to lower both barriers significantly, 
combining a finite value of the triaxiality angle $\gamma$ with finite octupole 
deformation near the outer barrier. 
 
We did not start the development of BSkG3 with the aim to further improve its 
description of the physics at large deformations and thus did not anticipate its
improved fission properties as compared to BSkG2. In an attempt to explain the differences, we investigated 
the surface energy and surface symmetry coefficients of both models through 
ETFSI calculations of semi-infinite nuclear matter along the lines of 
Ref.~\cite{Jodon16,Shchechilin20}. Both models have an essentially identical 
surface energy coefficient $a^{\rm ETF}_{\rm surf} = 17.5$ MeV\footnote{This value
differs from the value we quoted for BSkG2 in Ref.~\cite{Ryssens23}, 
$a^{\rm HF}_{\rm surf} = 17.9$ MeV. In Ref.~\cite{Ryssens23} however, we
relied on Hartree-Fock calculations of semi-infinite nuclear matter. 
Different approaches provide absolute values for $a_{\rm surf}$ that
can differ by up to 1~MeV, but the differences between parameterisations are 
robust across many-body methods, see the discussion in Ref.~\cite{Jodon16}. }
but have different surface symmetry coefficients $a^{\rm ETF}_{\rm s,surf}$:
they amount to -60.5 and -51.3 MeV for BSkG2 and BSkG3, respectively. These
values both fall within the (large) spread of predictions of other Skyrme-based
models~\cite{Nikolov11}.

\begin{figure}
  \includegraphics[width=0.45\textwidth]{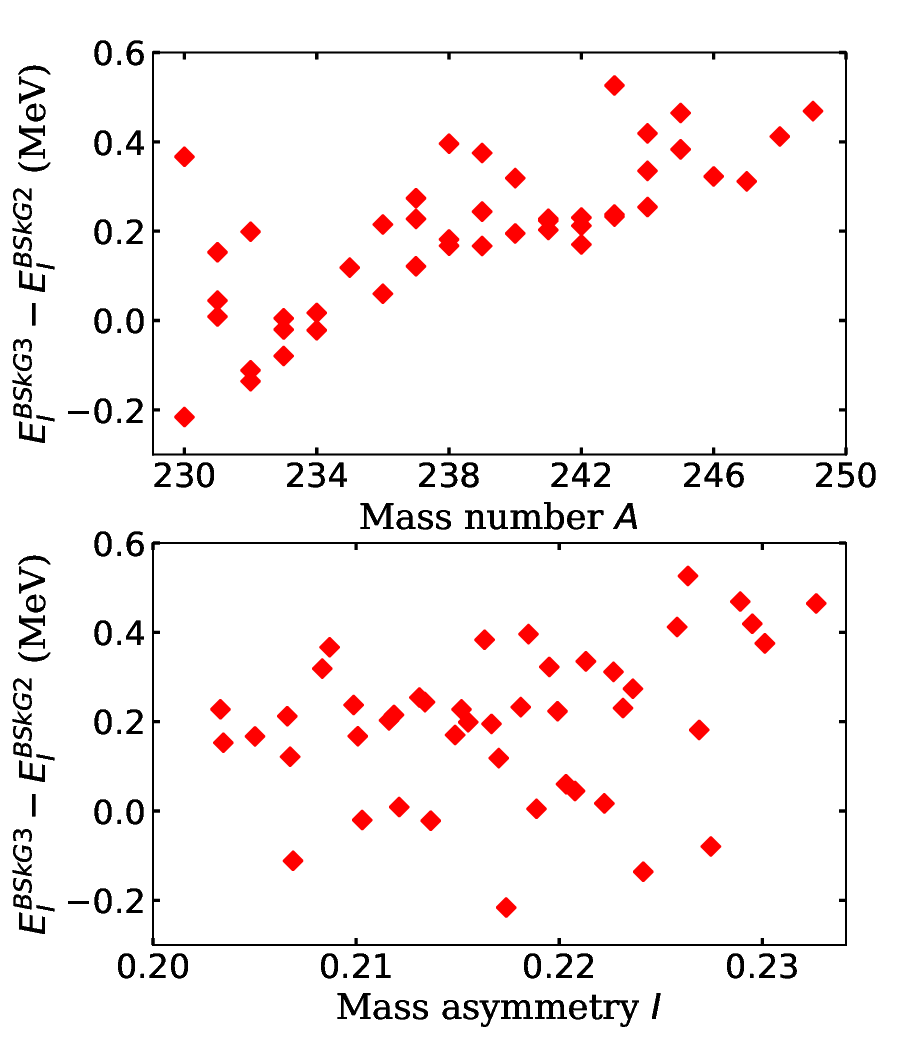}
\caption{Differences between calculated values for the primary fission barrier heights with BSkG3 and BSkG2 as a function of atomic mass $A$ (upper panel) or mass asymmetry $I$ (lower panel). }
\label{fig:evolution_bar}   
\end{figure}

We show the difference between the primary fission barriers obtained with BSkG3
and BSkG2 in Fig.~\ref{fig:evolution_bar}, both as a function of mass number 
$A$ (top panel) and asymmetry $I=(N-Z)/(N+Z)$ (bottom panel). This figure 
shows that the differences between both models are not large (less than 600
keV for any given nucleus) but that the predicted mass trends nevertheless
differ. Despite what one would naively expect from the difference in 
$a^{\rm ETF}_{\rm s,surf}$, there is no clear trend with asymmetry to be discerned
although this could simply be due to the limited variation of $I$ spanned by
the nuclei for which empirical barriers are available. 

Since BSkG3 differs from BSkG2 in many respects, we are unable to pinpoint a 
single source of the improved performance of the former but conclude that the 
differences arise from a multitude of (possibly competing) effects. One is simply
the inclusion of the entire set of 45 nuclei in the fitting protocol, as opposed to just 
the properties of twelve even-even nuclei as for BSkG2. The extended form of 
the EDF and the more microscopic treatment of pairing could also each
contribute to the differences between both models, as could the vibrational
correction which is significantly smaller in the new model. A detailed analysis
including nuclei outside of the actinide region is needed to disentangle these
effects but is beyond the scope of this work. For now, we only note that the
different mass trend and $a_{\rm s,surf}$ will likely result in varying  
predictions regarding the fission properties of extremely neutron-rich isotopes,
even if they produce modest variations for the fission barriers of actinide nuclei.

%
\section{Infinite nuclear matter properties}
\label{sec:INM}

We present in Tab.~\ref{tab:nucmatter} the properties of INM at saturation 
for BSk31\cite{Goriely16}, BSkG2\cite{Ryssens22} and BSkG3. We show, from top to
bottom, the Fermi momentum $k_{F}$, the saturation density $\rho_{\sat}$, the energy per particle of symmetric matter at saturation
$a_v$, the symmetry energy coefficient $J$, the slope of the symmetry energy 
$L$, the isoscalar (isovector) effective mass $M_s^*/M$ ($M_v^*/M$), the 
compressibility modulus $K_v$, the isovector component of the compressibility 
coefficient $K_{\rm sym}$, the skewness parameter $K^{\prime}$ and the 
Landau parameters $G_{0}$ and $G_0^{\prime}$ of symmetric INM.
Note that BSkG2 constrained the value of $J = 32$  MeV to obtain a minimum stiffness of the symmetry energy of the model.
For BSkG3, we fix $J = 31$ MeV, as previously done for BSk31, since we can achieve a stiff symmetry energy thanks to the extended Skyrme form and the new constraint at high density.
Among all parameters of Tab.~\ref{tab:nucmatter}, we note a major difference on $K_{\rm sym}$; BSkG2 presents a much lower value compared to BSkG3 and BSk31. 
This nuclear parameter still presents a large uncertainty. 
Different estimations, from unitary-gas considerations~\cite{Tews2017}, neutron skin of $^{48}$Ca and $^{208}$Pb~\cite{Sagawa2019},
and theoretical predictions based on the equilibrium density of nuclear matter lead to negative values $\approx - 100$ MeV ($\pm 100$ MeV)~\cite{Grams22c}.
Therefore, all three values are within the current uncertainty range. 
This parameter is obtained through the second 
derivative of the symmetry energy with respect to the density and evaluated at 
saturation. Since it is a high-order parameter of nuclear matter, 
$K_{\rm sym}$ is a measure of the stiffness of the symmetry energy at 
intermediate-high densities of a given model. 
The parameterisations BSk31 and 
BSkG3 were constrained to reproduce stiff NeuM EoS 
at high densities, as required by the existence of pulsars with $M > 2M_\odot$.
It is, therefore, natural that these models yield higher (less negative) values for $K_{\rm sym}$ than BSkG2.

It is interesting to note the differences in the INM parameters when constructing a model that reconciles an excellent reproduction of nuclear data with NS properties.
Values of the low-order nuclear parameters, as the first four lines of Tab.~\ref{tab:nucmatter}, are essential to the success of the mass table. In turn, the high-order ones, as $K_{\rm sym}$, play no role in the nuclear masses but strongly impact the high-density EoS and NS properties.
Note also that the impact of $K_{\rm sym}$ value is model dependent. 
A $K_{\rm sym} < - 100$ MeV (as obtained in BSkG2) does not necessary lead to incompatibility with NS data for all possible models. A recent study~\cite{Hana21} presents a set of EoS with $K_{\rm sym}$ ranging from $+20$ MeV to $-208$ MeV, which successfully reproduces the properties of NS.

\begin{table}[t!]
\centering
\caption{INM properties for the BSk31~\cite{Goriely16}, BSkG2~\cite{Ryssens22} and BSkG3 parameterisations. See Refs.~\cite{Goriely16,Margueron02,Chamel10} 
         for the various definitions. }
\tabcolsep=0.01cm
\begin{tabular}{l *{3}{d{6.5}} }
\hline\noalign{\smallskip}
Properties             & {\rm BSk31 } &  {\rm BSkG2 }   & {\rm BSkG3 }   \\
\hline
$k_F$~[fm]             &    1.3290      &    1.3265   & 1.3270  \\
$\rho_{\sat}$~[fm$^{-3}$] & 0.15855      &    0.15767   & 0.15784  \\
$a_v$~[MeV]            &  -16.110      &  -16.070   & -16.082 \\
$J$~[MeV]              &   31.000       &   32.000    & 31.000    \\
$L$~[MeV]              &   53.076        &  53.027     & 55.061 \\
$M_s^*/M$              &    0.84000      &    0.86000   & 0.85860 \\
$M_v^*/M$              &    0.73246      &    0.77330   & 0.72171 \\
$K_v$~[MeV]            &  244.01        &  237.45     & 242.56 \\
$K_{\text{sym}}$~[MeV] &  -15.780       & -150.60    & -21.248 \\
$K^\prime$~[MeV]       &  302.97        &  376.26     &  304.17 \\
$G_0$                  &    0.36551     &    0.35846    & 0.20046 \\
$G_0^\prime$           &    0.96946     &    0.97873    & 0.97812 \\
\hline
\end{tabular}
\label{tab:nucmatter} 
\end{table}

\begin{figure*}
\begin{center}
\includegraphics[width=0.8\textwidth]{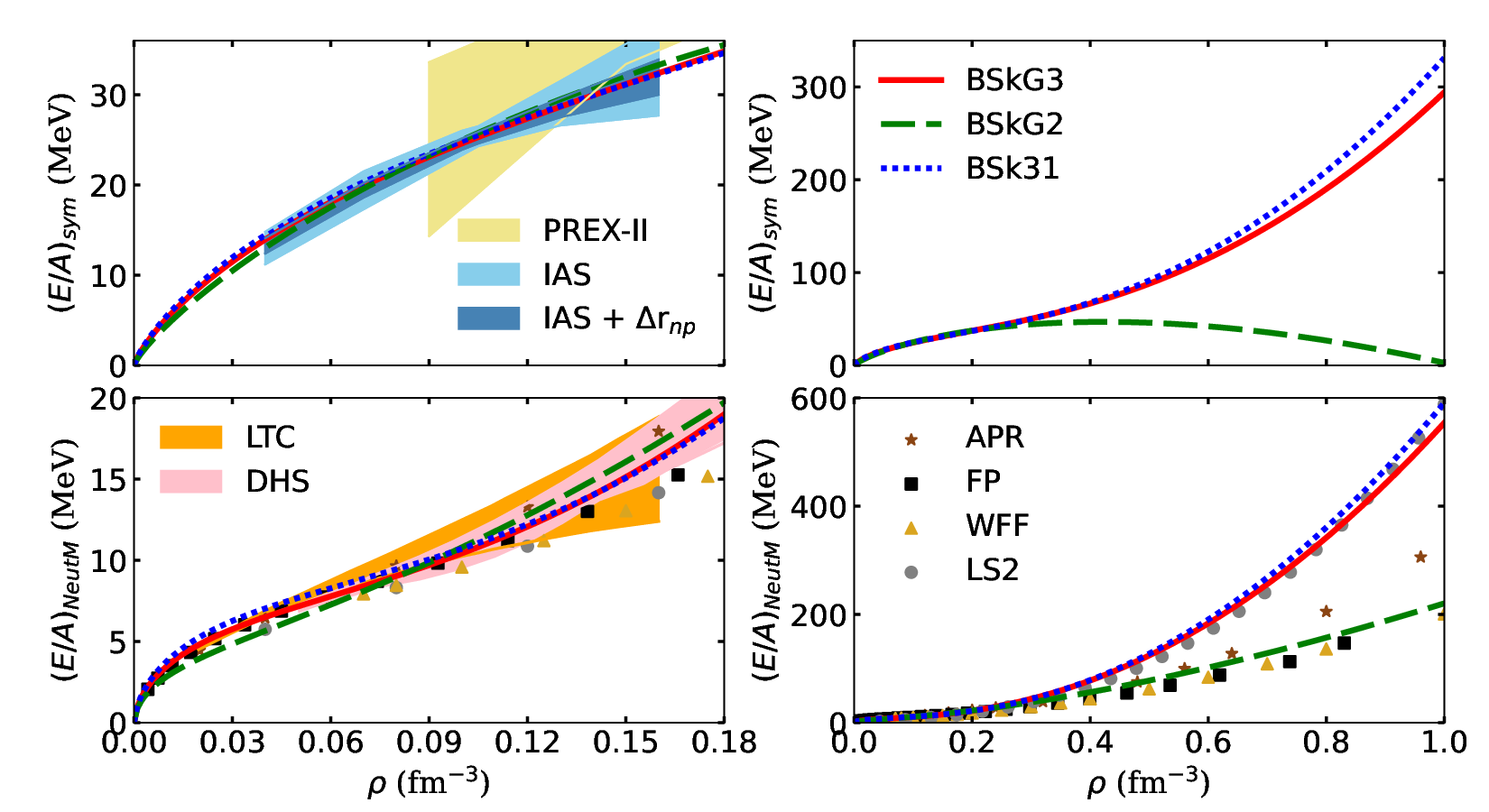}
\caption{Top (bottom) panels show the symmetry (neutron matter) energy per nucleon as a function of the baryon density for BSkG3 (continuous red), BSkG2 (dashed green), and BSk31 (dotted blue). Left panels show the predictions at densities below and around saturation while the right panels display the results at high densities. The models predictions are compared with experimental constraints~\cite{Danielewicz2014,Brendan2021} and ab-initio calculations of WFF \cite{WFF}, APR \cite{APR}, LS2 \cite{LS2}, FP \cite{FP},
LTC~\cite{Lynn16} and DHC~\cite{Drischler19}.}
\label{fig:eNeutMatt}  
\end{center}
\end{figure*}

We show in the bottom panels of Fig.~\ref{fig:eNeutMatt} the NeutM energy
 for BSkG3, BSkG2 and BSk31 and compare with the ab-initio calculations of WFF \cite{WFF}, APR \cite{APR}, LS2 \cite{LS2} and FP \cite{FP}.
 Chiral interactions predictions are represented by LTC~\cite{Lynn16} and DHC~\cite{Drischler19} bands.
At densities close to saturation and below (bottom left) all models present similar behavior, and are in good agreement with the ab-initio calculations. 
At high densities, however, the BSkG3 and BSk31 parameterisations present a much stiffer behaviour when compared to BSkG2 and are in good agreement with the LS2 EoS \cite{LS2}.
This figure highlights the main improvement in the BSkG3 INM properties compared to our previous BSkG1 
and BSkG2, which is the stiffness of the NeuM energy at high densities. 
The typical central density in NSs cores computed with Skyrme EDFs reaches 
6-10 times saturation density depending on the stiffness of the EoS.
However, EDFs fitted to experimental atomic masses 
only, remain poorly constrained at densities well above saturation density. The 
EoSs obtained with standard Skyrme EDFs are generally found to be too soft 
at high densities to explain the existence of the most massive observed NS. 
As shown, e.g., in Ref~\cite{Goriely10}, the EoS can be made stiff enough 
by introducing $t_4$ and $t_5$ terms without deteriorating the quality of 
the fit to atomic masses.
Top panels of Fig.~\ref{fig:eNeutMatt} display the symmetry energy for BSkG3, BSkG2, and BSk31. 
Note that we use the definition of symmetry energy as the energy necessary to convert symmetric nuclear matter (SNM) into pure NeutM: $e_{\rm sym} = e_{\rm NeutM} - e_{\rm SNM}$.
The top left panel compares the model predictions for $e_{\rm sym} $ with experiments constraints from the isobaric analog state (IAS) and IAS + neutron skin, $\Delta r_{np}$ (blue and dark blue), from Ref.~\cite{Danielewicz2014}
The yellow band shows the PREX-II predictions for the symmetry energy, where we vary $J = 38.1 \pm 4.7$ MeV and $L = 106 \pm 37$ MeV as suggested by Ref.~\cite{Brendan2021}. 
Note that all models reproduce well the experimental constraints given by the blue bands (IAS and IAS + $\Delta r_{np}$). However, models underestimate PREX-II at saturation and higher densities, being inside the yellow band just at low densities. This demonstrates again the tension raised by PREX-II results when compared with other experiments and theory predictions, as mentioned in Sec.~\ref{sec:bskg3param}.
The top right panel of Fig.~\ref{fig:eNeutMatt} shows the symmetry energy at high densities, where the effect of the stiff NeutM EoS can be observed avoiding the collapse of $e_{\rm sym} $ for BSkG3 and BSk31, in contrast to BSkG2.

\begin{figure}
\includegraphics[width=0.45\textwidth]{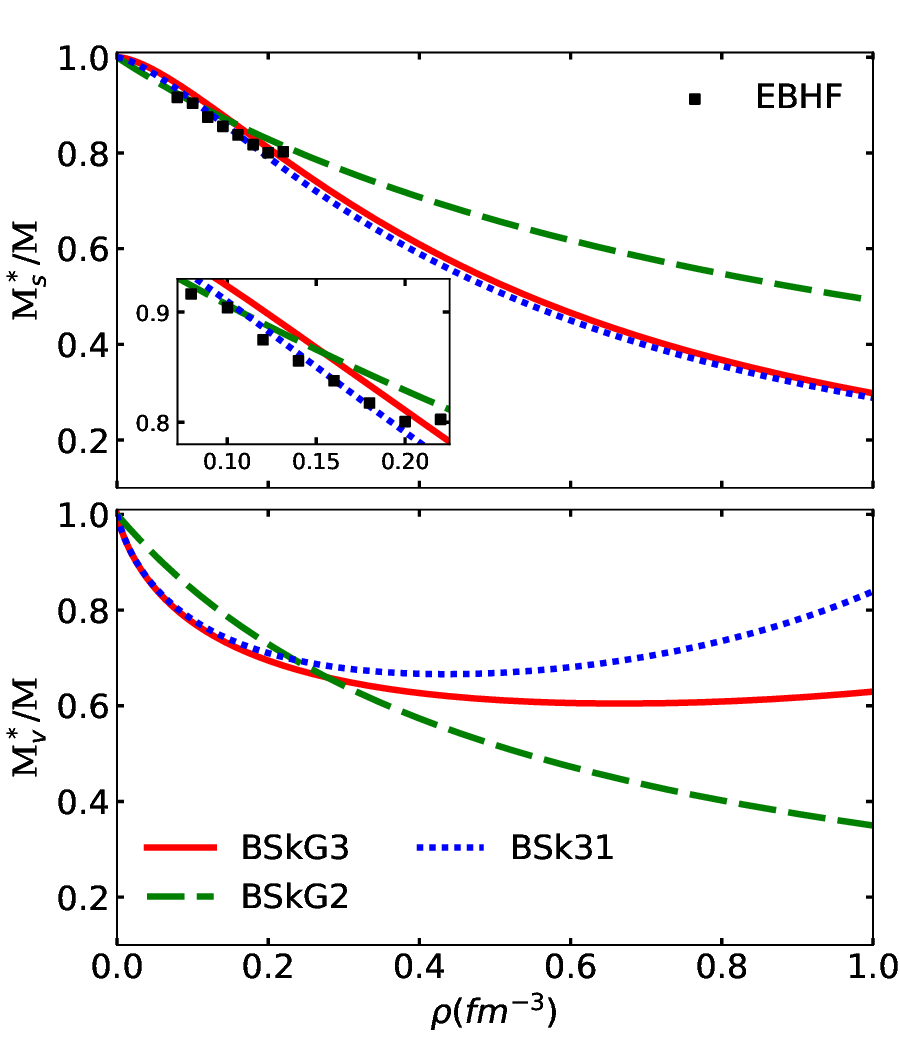}
\caption{Isoscalar (top) and isovector (bottom) effective mass as a function of the baryon density for BSkG3 (continuous red), BSkG2 (dashed green), and BSk31 (dotted blue). In the upper panel we compare the isoscalar effective mass with the EBHF calculations from Ref.~\cite{Cao06b}. 
}
\label{fig:msmv}  
\end{figure}

The density dependence of the effective mass is shown in Fig.~\ref{fig:msmv} for BSkG3 and compared with BSkG2 and BSk31. We note that for the isocalar channel, the effective mass obtained with BSkG3 is  almost identical to that of BSk31; it is close to that of BSkG2 up to $\rho \approx 0.2$ fm$^{-3}$ but is significantly smaller at higher  densities. 
Note that BSkG2, BSkG3 and BSk31 are in close agreement with the extended Brueckner-Hartree-Fock (EBHF) calculations of Ref.~\cite{Cao06b}, as shown in the insert plot in the top panel.
For the isovector effective mass, BSkG3 
is similar to BSk31, and lower than BSkG2, at lower densities. From densities around 0.3 fm$^{-3}$ and beyond its value lies in between BSkG2 and BSk31 predictions.
Moreover, BSkG3 shows $M_s^*/M = 0.859$ at saturation which is in good agreement with values obtained by EBHF calculations~\cite{Cao06b,Zuo02}.
Table~\ref{tab:nucmatter} shows that BSkG3 values for $M_s^*/M$ and $M_v^*/M$ taken at saturation present the hierarchy $M_s^* > M_v^*$, which implies that the neutron effective mass is higher then the proton one in neutron-rich matter. This hierarchy between neutron and proton effective mass is consistent with measurements of the isovector giant dipole resonance~\cite{Lesinski06} and ab-initio calculations~\cite{Cao06,Cao06b}. In particular, the BSkG3 magnitude of the splitting at saturation density is in excellent agreement with the realistic calculation of Ref.~\cite{Cao06b}.

\begin{figure*}
\begin{center}
\includegraphics[width=0.8\textwidth]{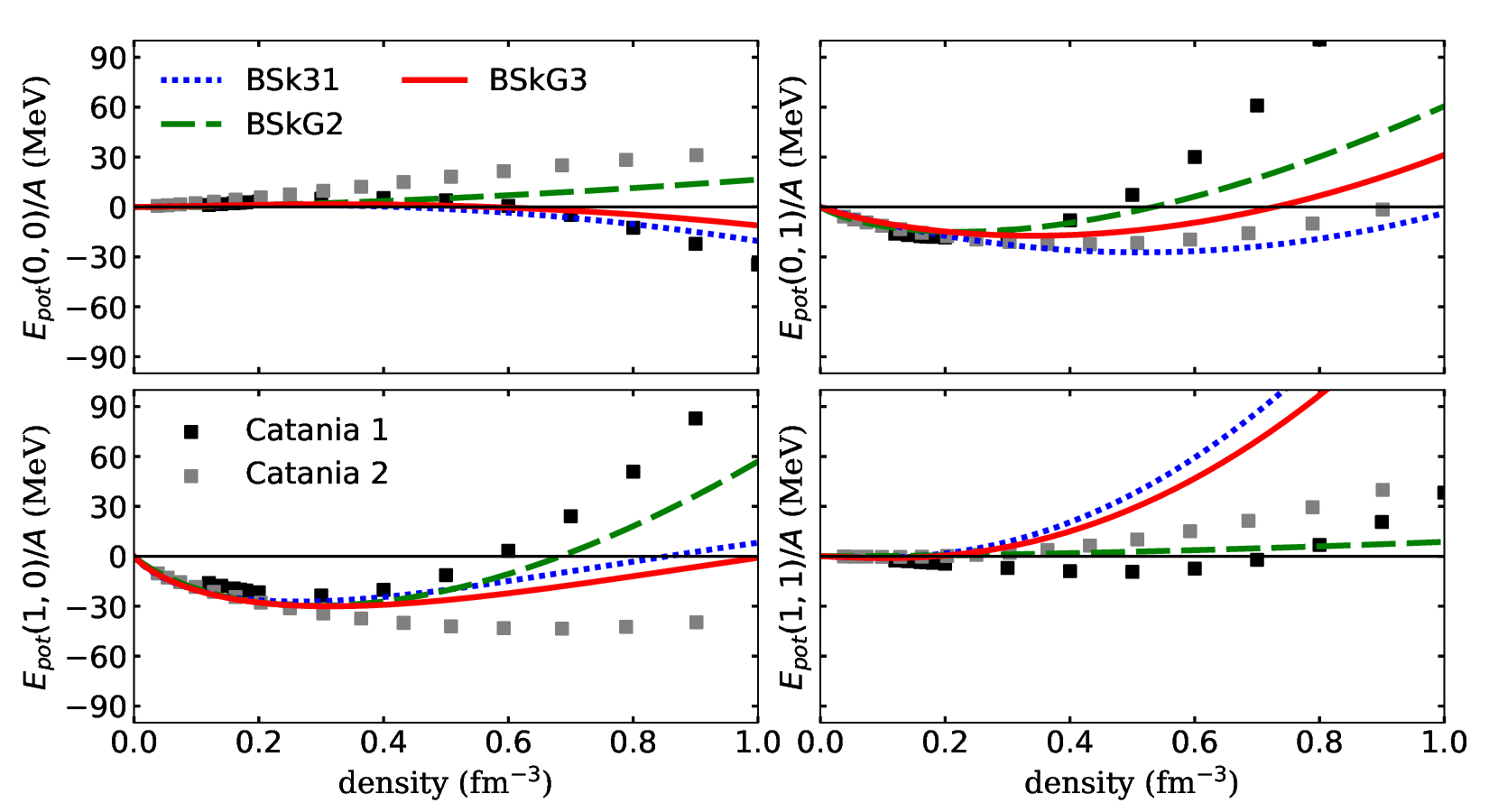}
\caption{Potential energy per particle in each ($S$,$T$) channel as a function of the baryon density for charge-symmetric nuclear matter for BSk31 (dotted blue), BSkG2 (dashed green) and BSkG3 (continuous red). Black and gray squares correspond to the "Catania 1"~\cite{catania1} and "Catania 2"~\cite{catania2} BHF calculations, respectively. }
\label{fig:chanels}  
\end{center}
\end{figure*}

We show in Fig.~\ref{fig:chanels} the distribution of the potential energy among the four two-body spin-isospin ($S, T$) channels for BSkG3 (continuous red), BSkG2 (dashed green) and BSk31 (dotted blue). We compare our results with two different Brueckner-Hartree-Fock (BHF) calculations labeled "Catania 1"~\cite{catania1} and "Catania 2"~\cite{catania2}. The model predictions for BSkG3 can be considered as satisfactory given the current uncertainty on what the real distribution actually is.


To summarize this section, we emphasise that BSkG3 exhibits INM properties 
at saturation that are similar to those of the previous BSkG models~\cite{Scamps21,Ryssens22} (see Tab.~\ref{tab:nucmatter}) 
with two important improvements: we now produce a stiff NeutM EoS at high 
densities (Fig.~\ref{fig:eNeutMatt}) which is important for 
explaining NSs with more than 2$M_\odot$, and replace the phenomenological 
pairing interaction of previous models by a more microscopically grounded 
interaction designed to match the $^1S_0$ pairing gaps in INM deduced from 
ab initio calculations (Fig.~\ref{fig:INM_gaps}).

\section{Neutron star properties}
\label{sec:NS}

\begin{figure}
  \includegraphics[width=0.47\textwidth]{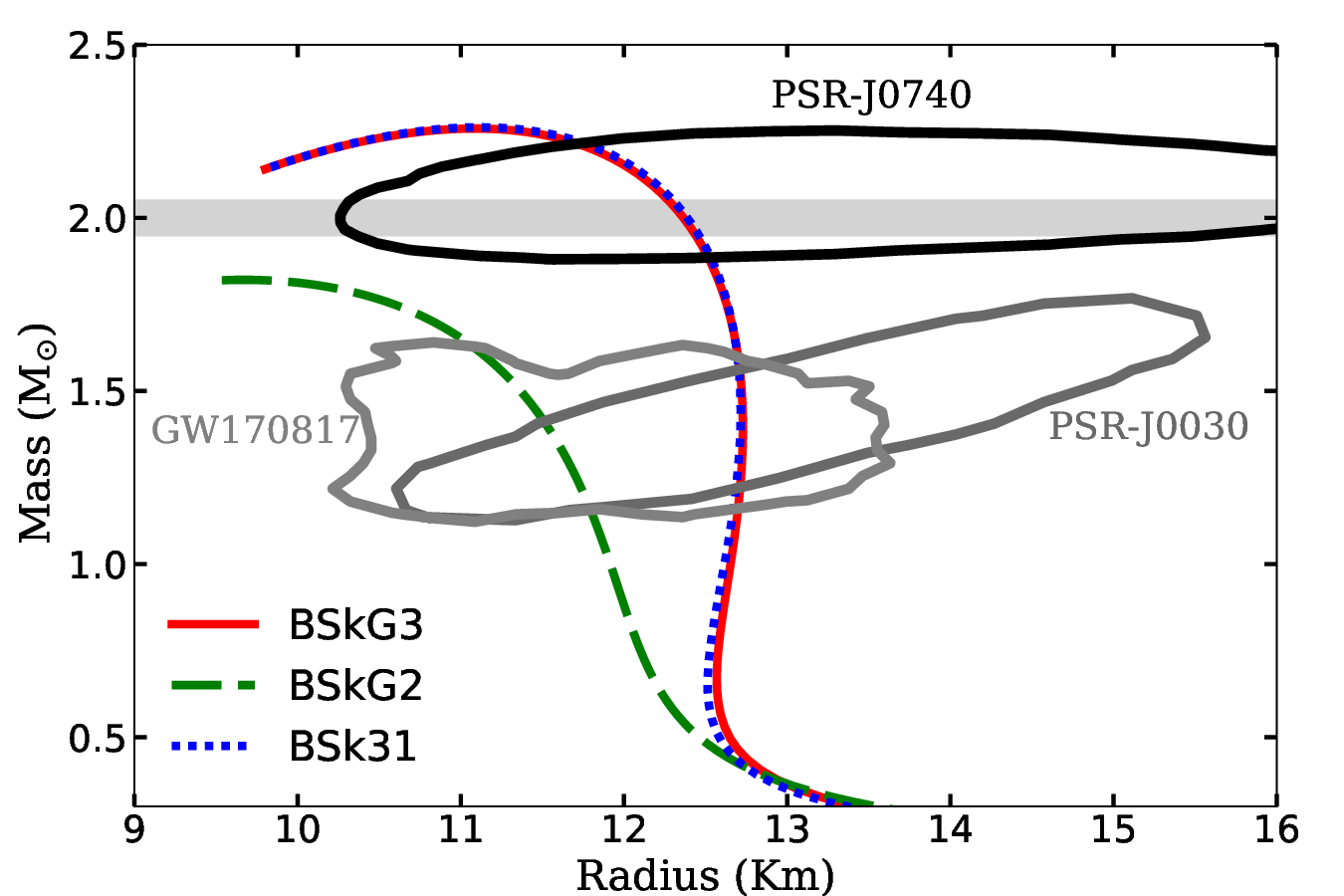}
\caption{Mass -- radius relation for BSkG3 (solid red line), BSkG2 (dashed green line) and BSk31 (dotted blue line). Contours show the NICER observations for the pulsars PSR-J0030~\cite{Riley19} and PSR-J0740~\cite{Miller21,NICER2021} together with LIGO/Virgo observation of GW170817~\cite{LIGOScientific:2018cki}. The gray band indicates the maximum mass constraint of 2M$_\odot$ NS~\cite{Demorest2010,Antoniadis2013}. See text for more details. }
\label{fig:tov}       
\end{figure}

In this section, the NS properties obtained with the new BSkG3 parameterisation are explored. 
We show in Fig.~\ref{fig:tov} the NS mass and radius relations for BSkG2, BSkG3, and BSk31.
We compare the model predictions with the observational data from X-ray emissions of NICER~\cite{Riley19,Miller21,NICER2021} and the GW170817 observation from the LIGO/Virgo interferometers~\cite{LIGOScientific:2018cki}.
The tidal deformability is strongly correlated to the NS mass and radius (see, e.g., Ref.~\cite{perot2019} and references therein), 
converting the uncertainty on $\tilde{\Lambda}$ into an uncertainty on the radius of a 1.4M$_\odot$ NS, as shown in the light gray contour in Fig.~\ref{fig:tov} indicated by ``GW170817''.

We note that BSkG3 and BSk31 models are compatible with all observational data while BSkG2 fails to reach the required maximum mass of 2$M_{\odot}$. This is a direct effect of the stiffness of the NeutM EoS, since the $\beta$-equilibrated matter contained in NSs is extremely neutron-rich. 
To obtain the NS macroscopic properties of Fig.~\ref{fig:tov}, we estimate the NSs masses and radii as follows. First, we  integrate the Tolman-Oppenheimer-Volkoff equations~\cite{Tolman1939,Oppenheimer1939} from the center of the star up to the crust-core transition fixed at the baryon number density $n_{\rm cc}=0.08$~fm$^{-3}$. The core is described using the corresponding EoS of neutron-proton-electron-muon ($npe\mu$) matter in $\beta$ equilibrium as in Ref.~\cite{Pearson18}. In this way, we obtain the mass $M_{\rm core}$ and radius $R_{\rm core}$ of the NS core. From the approximate formulas given in Ref.~\cite{Zdunik17}, we infer the NS mass and radius:  
\begin{equation}
M=M_{\rm core}+M_{\rm crust} \, , 
\end{equation}
\begin{equation}
R=R_{\rm core}\left\{1-\left[\left(\dfrac{\mu_{\rm cc}}{\mu_0} \right)^2 -1\right]\left(\dfrac{R_{\rm core} c^2}{2 G M} -1\right) \right\}^{-1} \, , 
\end{equation}
with 
\begin{eqnarray}
M_{\rm crust}=7.62\times 10^{-2} M_\odot \left(\dfrac{P_{\rm cc}}{\textrm{MeV~fm}^{-3}}\right) \notag \\ 
\times \left(1-\dfrac{2 G M_{\rm core}}{R_{\rm core}c^2} \right)\left(\dfrac{R_{\rm core}}{10~\textrm{km}}\right)^4 \left(\dfrac{M_\odot}{M_{\rm core}}\right)\, .
\end{eqnarray}
Here $\mu_{\rm cc}$ and $\mu_0$ denote the baryon chemical potentials at the crust-core transition and at the stellar surface, respectively. For cold catalyzed matter, the latter coincides with the mass per nucleon of $^{56}$Fe. $P_{\rm cc}$ represents the pressure at the crust-core transition. 

\begin{table}[t!]
\centering
\caption{NS properties obtained with BSk31, BSkG2 and BSkG3 parameterizations. We show in the first row the radius of the canonical NS of $1.4 M_\odot$; the following three lines give the mass, radius and energy density for the most compact NS. $\mathcal{E}_{\rm causal}$ denotes the energy density for which causality is violated. Next we show the threshold energy density and NS mass for the onset of the dUrca process and the energy density $\mathcal{E}_{\rm pQCD}$ up to which the model is consistent with constraints derived from pQCD (see text for details). }
\tabcolsep=0.01cm
\begin{tabular}{l *{3}{d{6.5}} }
\hline
Properties                  &   {\rm BSk31}  & {\rm BSkG2} & {\rm BSkG3}   \\
\hline
$R_{\rm 1.4}$~[km]                                 & 12.71   &  11.53  & 12.73  \\
$M_{\rm max}$~[$M_{\odot}$]                        & 2.26    & 1.82    & 2.26  \\
$R_{\rm M_{\rm max}}$~[km]                         & 11.11   & 9.66    & 11.10 \\
$\mathcal{E}_{\rm M_{\rm max}}$~[$10^{15}$ g cm$^{-3}$] & 2.276  & 3.192   &  2.273\\
$\mathcal{E}_{\rm causal}$~[$10^{15}$ g cm$^{-3}$]      & 2.392  & 6.558   & 2.381  \\
$\mathcal{E}_{\rm dUrca}$~[$10^{14}$ g cm$^{-3}$]  & 7.166   &   -     & 7.136  \\
$M_{\rm dUrca}$~[$M_{\odot}$]                      & 1.40    &   -     & 1.38  \\
$\mathcal{E}_{\rm pQCD}$~[$10^{15}$ g cm$^{-3}$]   & 2.509   & 4.289   & 2.508  \\
\hline
\end{tabular}
\label{tab:NS}
\end{table}

In Table~\ref{tab:NS}, we indicate the NS properties for BSk31, BSkG2, and BSkG3. The first three rows correspond, respectively, to the radius of the canonical 1.4 $M_\odot$ NS, the maximum mass of NS and its respective radius. 
We have also computed the energy density $\mathcal{E}_{\rm causal}$ of beta-equilibrated matter for which the sound speed becomes higher than the speed of light. We remark that $\mathcal{E}_{\rm causal}$ for BSkG3 is higher than the energy density $\mathcal{E}_{\rm M_{\rm max}}$ at the center of the most massive NS, as seen in Table~\ref{tab:NS}. Therefore none of the models presented here violates causality.

Table~\ref{tab:NS} also shows the density and the corresponding NS mass for which the proton fraction inside the star reaches the threshold for the onset of the direct Urca (dUrca) process~\cite{Lattimer91} for the three models. This process is required to interpret the observed thermal luminosity of some NSs~\cite{Burgio21}.
The model BSkG2 does not fulfill this constraint. 
BSk31 and BSkG3 models allow for the dUrca process for NSs with a mass above $\sim 1.4$ M$_{\odot}$.

Our models can be further tested using results from perturbative quantum chromodynamics (pQCD) calculations: 
assuming only causality and thermodynamic consistency, pQCD predictions at extreme 
densities ( $\gtrsim$ 40 $\rho_{\rm sat}$) impose constraints on the EoS
at lower densities~\cite{Komoltsev22}. We have verified that the 
BSkG3 EoS is consistent with this constraint up to an energy density 
$\epsilon_{\rm pQCD} = 2.508 \times 10^{15}$ [g cm$^3$], i.e. above 
the density of the most massive NS predicted by this model. 
BSk31 and BSkG2 also satisfy this constraint. 
\section{Conclusions and outlook}
\label{sec:conclusions}

We have presented a new entry in the BSkG-series of
EDF-based models. BSkG3 takes up the challenge of describing dense matter in 
a vast range of density regimes, combining a sophisticated description of thousands of 
atomic nuclei at saturation density in terms of symmetry-broken mean-field 
configurations with realistic predictions for dense matter at higher densities, such as encountered in NSs. The key to achieving both goals is the form of the 
functional: we move here to the extended form of Ref.~\cite{Chamel09} that includes 
additional density dependencies compared to the traditional Skyrme form. 

In contrast to its predecessors, the new model predicts an EoS of dense
matter that is entirely compatible with the most recent theoretical and 
observational NS constraints, including the existence of heavy pulsars with 
$M\geq 2 M_{\odot}$. We have also refined our approach to nucleon pairing, rendering the model more suited to the treatment of 
superfluidity in different NS layers by replacing the standard (and entirely 
phenomenological) ansatz for the pairing EDF with a more microscopically 
founded prescription designed to match the $^1S_0$ pairing gaps in INM as 
obtained from realistic EBHF calculations.

BSkG3 also describes nuclear properties with increased accuracy
   compared to BSkG2: we report rms deviations of 0.631 MeV on 2457 atomic 
   masses and 0.0237 fm for 810 charge radii. These global deviations remain 
   larger than the lowest ever achieved by the BSk models, but BSkG3 essentially
   matches the performance of the best of the older models for mass differences.
   In addition to this quantitative improvement, we have for the first time 
   accounted for reflection asymmetry in the nuclear ground state in even-even, 
   odd-mass and odd-odd nuclei alike. This further enriches the BSkG models
   in terms of collective effects and allows us to connect to the phenomenology 
   of octupole deformation. Although the impact of this degree of freedom on 
   the global accuracy of our model is small due to the limited number of nuclei
   impacted, we anticipate meaningful effects for nucleosynthesis simulations
   chiefly because of the large impact of octupole deformation on the ground
   states of exotic neutron-rich  nuclei near $N \sim 190$.
   The accuracy of the new model extends beyond ground states to fission properties: BSkG3 
   reproduces the 45 RIPL-3 reference values for the primary barriers of 
   actinide nuclei with a rms deviation of 0.33 MeV, combined with rms deviations
   on secondary barriers and isomeric excitation energies of 0.51 MeV and 
   0.34 MeV respectively. These values are, to the best of our knowledge, 
   unmatched in the available literature.

This work addresses the extrapolation of the BSkG to high nucleon densities, 
   but this is hardly the only extrapolation relevant to NS physics: a 
   general-purpose EoS should not be limited to zero temperature but cover a range of temperatures, since 
   the temperatures involved in core-collapse supernovae and NS mergers simulations can 
   reach above 100 MeV~\cite{Oertel17,Perego2019}. To enlarge the reach of the BSkG models, 
   we plan to extend our fitting protocol to also include constraints on the properties
   of the EoS at finite temperature, as predicted by advanced many-body 
   approaches such as for example Refs.~\cite{Lu20,Shang20,Keller21,Carbone19}.

We have several prospects to further improve our description of atomic nuclei. 
   The accuracy we achieve here for fission barriers in the actinide region
   motivates us to extend our fission calculations to exotic neutron-rich nuclei.
   Although this is a monumental task because
   of the number of nuclei involved, efforts in this direction are under way.
   We also anticipate studying the issues touched upon in Sec.~\ref{sec:BSk_comparison}:
   like all EDF-based models, the BSkG-series do not describe a small but 
   systematic contribution to the binding energy of odd-odd nuclei,  
   contaminating all mass differences that involve such isotopes~\cite{Hukkanen22}.
   As suggested in Ref.~\cite{Ryssens22}, tuning the proton-neutron spin-spin 
   term in the time-odd part of the functional might provide a way forward.
   Our most ambitious and undoubtedly long-range goal is to improve our 
   description of nuclear collective motion, abandoning simple recipes such as 
   the rotational correction in favor for a more microscopic approach founded in
   beyond-mean-field techniques.


\begin{acknowledgements}
We gratefully acknowledge useful discussions with both Dr. M. Bender and 
N. Shchechilin on the subject of surface tension and its isospin dependency.
This work was supported by the Fonds de la Recherche Scientifique (F.R.S.-FNRS) and the Fonds Wetenschappelijk Onderzoek - Vlaanderen (FWO) under the EOS Projects nr O022818F and O000422F. 
The present research benefited from computational resources made available on the Tier-1 supercomputers Zenobe and Lucia of the Fédération Wallonie-Bruxelles,
infrastructure funded by the Walloon Region under the grant agreement nr 1117545.
Further computational resources have been provided by the clusters Consortium des Équipements de Calcul Intensif (CÉCI), funded by F.R.S.-FNRS under Grant No. 2.5020.11 and by the Walloon Region. 
G.S. is supported by U.S. Department of Energy, Office of Science, Grant No.DE-AC05-00OR22725. W.R. and S.G. gratefully acknowledge support by the F.R.S.-FNRS.
\end{acknowledgements}

\appendix

\section{List of abbreviations}
\label{app:acronyms}

\begin{tabular}{ll}
Abbreviation & Meaning \\
\hline
BSkG  & Brussels-Skyrme-on-a-Grid \\
dUrca & direct Urca (process) \\
EDF   & Energy density functional \\
EoS   & Equation of state \\
INM   & Infinite (homogeneous) nuclear matter        \\
MLNN  & Multilayer neural network \\
MOI   & Moment of inertia \\
NeutM & Pure (infinite) neutron matter \\
$npe\mu$ & neutron-proton-electron-muon \\
NS    & Neutron star  \\
SNM  & symmetric (infinite) nuclear matter \\
PES   & Potential energy surface  \\
pQCD  & Perturbative quantum chromodynamics \\
rms   & root-mean-square \\
r-process & rapid neutron-capture process \\
\hline
\end{tabular}

\section{Coupling constants of $E_{\rm Sk}$}
\label{app:couplingconstants}

We include here the complete expressions for all coupling constants of the 
Skyrme energy densities in terms of the model parameters. The coupling constants appearing
in the time-even, central part of the EDF are:
\begin{alignat}{2}
C^{\rho \rho}_t (\rho_0)        =& + C^+_{0t}(t_0,x_0)  + \frac{1}{6} C^+_{0t}(t_3,x_3) \rho_0^{\alpha} \, , \\
C^{\rho \tau}_t (\rho_0)        =& + \frac{1}{2} C^+_{0t}(t_1,x_1)  + \frac{1}{2} C^-_{0t}(t_2,x_2)                            \nonumber \\
                                 & + \frac{1}{2}C_{0t}^+(t_4,x_4)\rho_0^{\beta}  + \frac{1}{2}C_{0t}^-(t_5,x_5)\rho_0^{\gamma}  \, ,     \\  
C^{\rho \Delta \rho}_t          =& - \frac{3}{8} C^+_{0t}(t_1,x_1) + \frac{1}{8} C^-_{0t}(t_2,x_2)                                                 \, , \\
C^{\nabla \rho\nabla \rho}_t (\rho_0) =& + \frac{3}{8} C^+_{0t}(t_4,x_4)\rho_0^{\beta}
                                         - \frac{1}{8} C^-_{0t}(t_5,x_5) \rho_0^{\gamma}  \, , \\
C^{\rho \nabla \rho\nabla \rho}_t (\rho_0) =& - \frac{1}{2} C^+_{0t}(t_4,x_4) \rho_0^{\beta-1} \, ,
\end{alignat} 
while those in the time-odd, central part $\mathcal{E}_{t, \rm e}(\bold{r})$ are
\begin{alignat}{2}
C^{ss}_t (\rho_0)        =& C^+_{t, 11}(t_0, x_0)  + \frac{1}{6} C^+_{1t}(t_3,x_3) \rho_0^{\alpha} \, , \\
C^{jj}_t (\rho_0)        =& - C^{\rho \tau}_t(\rho_0) \, .
\end{alignat} 
We again draw the readers attention to a change of notation with respect
to the BSkG1 and BSkG2 models: the exponent of the density in the expression 
for $C^{\rho \rho}_t(\rho_0)$ was denoted by $\gamma$ in Refs.~\cite{Scamps21,Ryssens22}, 
whereas here we indicate
the same quantity here with $\alpha$ following Ref.~\cite{Goriely09}.

We define the shorthands $C^{+/-}_{ST}(t,x)$ as in Ref.~\cite{Ryssens21}:
\begin{alignat}{2}
C^{+}_{00} (t,x) &= +\tfrac{3}{8} t \, , \phantom{+\tfrac{1}{4} tx } & \quad
C^{+}_{01} (t,x) &= -\tfrac{1}{8} t - \tfrac{1}{4} tx           \, , \nonumber \\
C^{+}_{10} (t,x) &= -\tfrac{1}{8} t + \tfrac{1}{4} tx           \, , &
C^{+}_{11} (t,x) &= -\tfrac{1}{8} t \, , \phantom{+ \tfrac{1}{4} tx} \nonumber \\
C^{-}_{00} (t,x) &= +\tfrac{5}{8} t + \tfrac{1}{2} tx           \, , &
C^{-}_{01} (t,x) &= +\tfrac{1}{8} t + \tfrac{1}{4} tx           \, , \nonumber \\
C^{-}_{10} (t,x) &= +\tfrac{1}{8} t + \tfrac{1}{4} tx           \, , &
C^{-}_{11} (t,x) &= +\tfrac{1}{8} t \, .\phantom{+ \tfrac{1}{4} tx } 
\label{eq:CCshort}
\end{alignat}
Finally, the coupling constants of the time-even spin-orbit part of the EDF are
given in terms of the parameters $W_0$ and $W_0'$
\begin{alignat}{2}
C^{\rho \nabla \cdot J}_0    &= - \frac{W_0}{2} - \frac{W_0'}{4} \, , & \quad
C^{\rho \nabla \cdot J}_1    &= - \frac{W_0'}{4} \, . 
\end{alignat}
The time-odd spin-orbit coupling constants are identical to the time-even ones, 
i.e.~$ C^{j \nabla s}_t =C^{\rho \nabla \cdot J}_t$.

\section{Rotational, vibrational and Wigner energies}
\label{app:corrEnergies}

For completeness, we briefly recall here the expressions for the 
rotational, vibrational, and Wigner energies as employed in the BSkG1, BSkG2 and BSkG3 models.

\begin{itemize}
    \item {\it The rotational correction}
\end{itemize}

The rotational correction is based on a simple perturbative cranking model, involving the 
Belyaev moments of inertia (MOI) around 
the three principal axes of the nucleus,
$\mathcal{I}^{\rm B} _{\mu}$($\mu=x,y,z$)~\cite{Belyaev61}:
\begin{subequations}
\begin{align}
\label{eq:Erot}
E_{\rm rot} &= - 
\sum_{\mu=x,y,z} f_{\mu}^{\rm rot}\frac{\langle \hat{J}^{2}_{\mu} \rangle}{2 \mathcal{I}^{\rm B}_{\mu}} \, , \\
f_{\mu}^{\rm rot} &=  b  \tanh \left( c \frac{\mathcal{I}^{\rm B}_{\mu}}{\mathcal{I}_c}\right) \, ,
\label{eq:rotcut}
\end{align}
\end{subequations}
where $\hat{J}_{\mu}$ is an angular momentum operator and 
$\mathcal{I}_{\rm C} = \tfrac{2}{15} m R^2 A$ is (one-third of) the MOI of a rigid rotor of radius $R = 1.2 A^{1/3}$, comprised of $A$ nucleons of average mass $m$. Both $b$ and $c$ are adjustable parameters, while the three MOI in Eq.~\eqref{eq:Erot} are calculated consistently from the HFB auxiliary state.

\begin{itemize}
    \item {\it The vibrational correction}
\end{itemize}

The vibrational correction is given by:
\begin{subequations}
\begin{align}
\label{eq:vibrationalcorrection}
E_{\rm vib} &= 
-\sum_{\mu = x,y,z} f_{\mu}^{\rm vib}  \frac{\langle \hat{J}_{\mu}^2 \rangle}{2 \mathcal{I}^{\rm B}_{\mu}} \, , \\
f_{\mu}^{\rm vib} &= d B_{\mu} e^{- l \left( B_{\mu} - B_0 \right)^2} \, ,\\
B_{\mu} &= \frac{\mathcal{I}^{\rm B}_{\mu}}{\mathcal{I}_c} \, .
\label{eq:B_compare}
\end{align}
\end{subequations}
where $d, l$ and $B_0$ are model parameters. Note that Eq.~\eqref{eq:vibrationalcorrection} only intends to capture the deformation dependence of the vibrational energy, as discussed in Ref.~\cite{Goriely07,Ryssens22}.
\begin{itemize}
    \item {\it The Wigner energy}
\end{itemize}

Mean-field models that describe the nucleus with separate proton and neutron 
   treatments typically underestimate the binding energy of  $N \simeq Z$ nuclei. 
   To rectify this issue, the BSkG models include the Wigner energy correction
   of Ref.~\cite{Goriely2002}:
\begin{align}
E_{\rm W}=& V_W\exp(-\lambda((N-Z)/A)^2) \nonumber \\
         &+ V_W^{\prime}|N-Z|\exp\left[-(A/A_0)^2\right] \, ,
\label{eq:wig}
\end{align}
with four adjustable parameters $V_W, V_W', \lambda$, and $A_0$.

\section{Explanation of the supplementary material}
\label{app:explanation}

We provide as supplementary material the files \newline 
\textsf{Mass\_Table\_BSkG3.dat} and \textsf{Fission\_Table\_BSkG3.dat}.
The former contains the calculated ground state properties of all nuclei with 
$ 8 \leq Z \leq 118$ lying between the proton and neutron drip lines. The latter
contains the fission barriers and isomer excitation energies as calculated for 
all 45 nuclei with $Z \geq 90$ that figure in the RIPL-3 database. The contents 
of both files follow the conventions of the supplementary files of 
Refs.~\cite{Scamps21,Ryssens22} and \cite{Ryssens23} with only two exceptions.
First, we now also list the octupole deformations $\beta_{30}$ and $\beta_{32}$ 
among ground state properties. Second, for odd-mass and odd-odd nuclei with 
finite octupole deformation we can no longer assign a parity quantum number 
to the calculated ground states; in such cases we set the final two columns of
\textsf{Mass\_Table\_BSkG3.dat} that contain the parities of the blocked
quasiparticle(s) to zero. For convenience, we repeat the contents of all columns of both files in 
Tables~\ref{tab:suppl1} and \ref{tab:suppl2}.

\begin{table*}[]
\centering
\begin{tabular}{llll}
\hline
\hline
Column &  Quantity & Units & Explanation \\
\hline
1 & Z & $-$ & Proton number\\
2 & N & $-$ & Neutron number\\
3 & $M_{\rm exp}$ & MeV  & Experimental atomic mass excess \\
4 & $M_{\rm th}$ & MeV   & BSkG3 atomic mass excess \\
5 & $\Delta M$ & MeV      & $M_{\rm exp} - M_{\rm th}$\\
6 & $E_{\rm tot}$ & MeV   & Total energy, Eq.~\eqref{eq:Etot} \\
7 & $\beta_{20}$ & $-$    & \multirow{3}*{Quadrupole deformation} \\
8 & $\beta_{22}$ & $-$    & \\
9 & $\beta_2$  & $-$        & \\
10& $\beta_{30}$ & $-$    & \multirow{2}*{Octupole deformation} \\
11& $\beta_{32}$ & $-$    & \\
12 & $E_{\rm rot}$ & MeV  & Rotational correction \\
13 & $\langle \Delta \rangle_n$ & MeV & Average neutron gap\\
14 & $\langle \Delta \rangle_p$ & MeV & Average proton gap\\
15 & $r_{\rm BSkG3}$ & fm  & Calculated rms charge radius\\
16 & $r_{\rm exp}$  & fm  & Experimental rms charge radius\\
17 &$\Delta r$ & fm   &  $r_{\rm exp}- r_{\rm BSkG3}$  \\
18 & $\mathcal{I}^B$ & $\hbar^2$ MeV$^{-1}$ & Calculated Belyaev MOI.\\
19 & par(p) & $-$ & Parity of proton qp. excitation\\
20 & par(n) & $-$ & Parity of neutron qp. excitation\\
\hline 
\hline
\end{tabular}
\caption{Contents of the \textsf{Mass\_Table\_BSkG3.dat} file.}
\label{tab:suppl1}
\end{table*}

\begin{table*}[]
\centering
\begin{tabular}{lllll}
\hline
\hline
Column &  Quantity & Fission property & Units & Explanation   \\
\hline
1 & Z & & $-$ & Proton number \\
2 & N & & $-$ & Neutron number \\
\hline
3 & $E$          & Inner barrier  & MeV  & Barrier height          \\
4 & $\beta_{20}$ &                & $-$  & Quadrupole deformation  \\
5 & $\beta_{22}$ &                & $-$  &                         \\
\hline
6 & $E$          & Outer barrier  & MeV & Barrier height           \\
7 & $\beta_{20}$ &                & $-$ & Quadrupole deformation   \\
8 & $\beta_{22}$ &                & $-$ & \\
9 & $\beta_{30}$ &                & $-$ & Octupole deformation \\
\hline
10 & $E$         & Isomer         & MeV &  Excitation energy     \\
11 & $\beta_{20}$&                & $-$ &  Quadrupole deformation    \\
\hline 
\hline
\end{tabular}
\caption{Contents of the \textsf{Fission\_Table\_BSkG3.dat} file.}
\label{tab:suppl2}
\end{table*}

\bibliographystyle{spphys}       
\bibliography{ref}   

\end{document}